\documentclass[conference]{IEEEtran}
\IEEEoverridecommandlockouts
\usepackage{amsmath,amssymb,amsfonts}
\usepackage{algorithmic}
\usepackage{graphicx}
\usepackage{textcomp}
\usepackage{xcolor}
\usepackage{enumitem}
\usepackage[backend=biber, style=ieee]{biblatex}
\usepackage[hidelinks]{hyperref}
\usepackage{textcase}

\begin{document}

\title{AMR-MoEGA: Antimicrobial Resistance Prediction using Mixture of Experts and Genetic Algorithms}

\author{\IEEEauthorblockN{Anshul Bagaria}
\IEEEauthorblockA{\text{Department of Data Science and Artificial Intelligence}\\{Indian Institute of Technology Madras, India}}}
\maketitle

\begin{abstract}
Antimicrobial resistance (AMR) poses a mounting global health crisis, requiring rapid and reliable prediction frameworks that capture its complex evolutionary dynamics. Traditional antimicrobial susceptibility testing (AST), while accurate, remains laborious and time-consuming, limiting its clinical scalability. Existing computational approaches, primarily reliant on single nucleotide polymorphism (SNP)-based analysis, fail to account for evolutionary drivers such as horizontal gene transfer (HGT) and genome-level interactions.

This study introduces a novel Evolutionary Mixture of Experts (Evo-MoE) framework that integrates genomic sequence analysis, machine learning, and evolutionary algorithms to model and predict AMR evolution. A Mixture of Experts model, trained on labeled genomic data for multiple antibiotics, serves as the predictive core, estimating the likelihood of resistance for each genome. This model is embedded as a fitness function within a Genetic Algorithm designed to simulate AMR development across generations. Each genome, encoded as an individual in the population, undergoes mutation, crossover, and selection guided by predicted resistance probabilities.

The resulting evolutionary trajectories reveal dynamic pathways of resistance acquisition, offering mechanistic insights into genomic evolution under selective antibiotic pressure. Sensitivity analysis of mutation rates and selection pressures demonstrates the model’s robustness and biological plausibility. Validation against curated AMR databases and literature evidence further substantiates the framework’s predictive fidelity.

This integrative approach bridges genomic prediction and evolutionary simulation, offering a powerful tool for understanding and anticipating AMR dynamics, and potentially guiding rational antibiotic design and policy interventions.
\end{abstract}


\section{Introduction}
\label{sec:introduction}
The dawn of the antibiotic era, marked by the seminal discovery of \textit{Penicillium chrysogenum} by Alexander Fleming in 1928, ushered in a transformative period in modern medicine and healthcare. Antibiotics quickly became one of the most significant medical breakthroughs of the twentieth century. They dramatically reduced the morbidity and mortality linked to bacterial infections \autocite{Muteeb2023-bl}. Furthermore, they enabled previously high-risk medical procedures, including organ transplants, chemotherapy, and complex surgeries \autocite{Hutchings2019-du}. However, the widespread and often indiscriminate use of antibiotics has inadvertently precipitated one of the greatest challenges to global health today: the rise of \textbf{antimicrobial resistance (AMR)}. This phenomenon, now recognized as a silent pandemic, threatens to nullify decades of progress in infectious disease control \autocite{Dhingra2020-tq}, \autocite{Salam2023-ye}.

The emergence and propagation of AMR constitute a multifaceted global crisis that transcends public health, impinging upon agriculture, food security, and socioeconomic stability. Resistant pathogens continue to evolve and disseminate across ecological and geographical boundaries, giving rise to multidrug-resistant (MDR) and extensively drug-resistant (XDR) strains that defy existing therapeutic regimens \autocite{AHMED2024100081}, \autocite{Naylor2018-wz}. The \textbf{World Health Organization (WHO)} projects that by 2050, AMR could cause up to 10 million deaths annually and impose an economic burden exceeding USD 100 trillion if unchecked \autocite{Naylor2018-wz}. This looming catastrophe underscores the urgent necessity of developing rapid, scalable, and predictive methodologies capable of anticipating resistance evolution and guiding the design of next-generation therapeutics.

Conventional approaches for AMR analysis rely primarily on in vitro \textbf{antimicrobial susceptibility testing (AST)} and the determination of the \textbf{minimum inhibitory concentration (MIC)}. While these phenotypic assays are highly reliable, they are constrained by their dependence on cultivable isolates \autocite{Abushaheen2020-kr}, specialized laboratory infrastructure, and the time required for microbial growth—often spanning 24 to 72 hours \autocite{Gajic2022-kl}, \autocite{Boolchandani2019-cq}, \autocite{Barnes2023-ov}. Such delays can be fatal in acute infections where timely intervention is critical. Furthermore, these assays provide limited insight into the underlying genetic and evolutionary mechanisms driving resistance, thereby impeding proactive surveillance and intervention strategies.

In recent years, the convergence of genomics and machine learning has enabled the emergence of \textit{in silico} approaches for AMR prediction. High-throughput sequencing technologies, coupled with advances in computational biology, have facilitated genome-wide association analyses linking \textbf{specific single nucleotide polymorphisms (SNPs)} or resistance genes to phenotypic resistance profiles. For example, curated resources such as the \textit{Comprehensive Antibiotic Resistance Database (CARD)} now supports machine learning pipelines for resistance prediction and resistome analysis \autocite{Alcock2023-gh}, \autocite{Kim2022-fi}. Other tools such as \textit{ResFinder}, \textit{DeepARG}, and \textit{AMRFinderPlus} have successfully leveraged sequence-based features to predict antibiotic resistance genes from metagenomic data with increasing accuracy \autocite{Florensa2022-jf, Arango-Argoty2018-ej, Feldgarden2021-yl}. Despite these advances, most existing models remain static, i.e., they remain focused on the identification of resistance determinants at a single point in time. These have thereby overlooked the dynamic evolutionary processes such as \textbf{horizontal gene transfer (HGT)}, recombination, and adaptive mutations that continually reshape bacterial genomes \autocite{Kim2022-fi}, \autocite{chindelevitch2022applyingdatatechnologiescombat}.

Evolution, inherently stochastic and multi-dimensional, governs the emergence of resistance traits within microbial populations. Mechanisms like HGT enable the rapid dissemination of resistance-conferring genes across species boundaries, while selective pressures from antibiotic exposure drive the fixation of beneficial mutations \autocite{10.3389/fmicb.2019.01933}, \autocite{Sun2023-bo}. Understanding these evolutionary trajectories is thus pivotal to developing predictive models that not only classify resistance but also simulate its progression \autocite{Baquero2021-zm}, \autocite{Wilson2016-ap}. Traditional ML approaches lack the capacity to capture temporal adaptation or the non-linear fitness landscapes associated with microbial evolution \autocite{Hasan2021-pm}.

To address these limitations, this study proposes a novel \textbf{Evolutionary Mixture of Experts (EvoMoE)} framework that integrates genomic learning with evolutionary simulation for enhanced AMR prediction and analysis. The framework combines a \textit{Mixture of Experts (MoE)} model, trained on annotated genomic sequences for multiple antibiotics, with a \textit{Genetic Algorithm (GA)} designed to emulate evolutionary dynamics under antibiotic selection pressure. Each genome is represented as an individual in the GA population, characterized by its genomic features and an MoE-predicted probability of resistance. Evolutionary operators such as mutation, crossover, and selection are applied iteratively, guided by the MoE-derived fitness landscape, thereby simulating the adaptive progression of resistance over successive generations.

This integrative approach bridges the gap between static genomic prediction and dynamic evolutionary modeling. By tracing \textit{in silico evolutionary trajectories} and analyzing shifts in predicted AMR probabilities, the proposed framework provides mechanistic insights into the evolutionary pathways through which resistance may arise and proliferate. Sensitivity analyses across varying mutation rates and selection pressures further elucidate the robustness of these simulated pathways, while validation against literature-supported evolutionary patterns ensures biological plausibility.

Ultimately, the EvoMoE framework represents a big step towards predictive microbiology—offering a computational lens through which we can anticipate, rather than merely detect, the emergence of antimicrobial resistance. Such predictive capability could inform targeted surveillance, guide rational drug design, and support the development of adaptive therapeutic strategies to combat the escalating AMR crisis

\subsection{Related Work}
\label{subsec:related}
The growing urgency to combat AMR has catalyzed extensive research across multiple domains, including \textit{genomics}, \textit{computational biology}, and \textit{artificial intelligence}. Early studies predominantly focused on genomic variant analysis, identifying resistance-associated mutations through alignment-based and statistical methods. With the advent of high-throughput sequencing, the scope of AMR prediction expanded to include machine learning and deep learning approaches capable of integrating large-scale genomic, transcriptomic, and proteomic datasets. Recent efforts have explored the incorporation of HGT information, plasmid dynamics, and pan-genomic architectures to capture the complex evolutionary interplay driving resistance dissemination. Simultaneously, evolutionary modeling and algorithmic simulations, ranging from agent-based models to genetic algorithms, have been employed to mimic microbial adaptation under antibiotic pressure. However, these domains often operate in isolation, with predictive models lacking evolutionary realism and evolutionary simulations neglecting fine-grained genomic determinants of resistance.

\subsubsection{\textbf{Genomic Machine Learning-Based AMR Prediction}}
\label{subsubsec:genomic_ml}
Advances in high-throughput sequencing have enabled the transition from phenotype-based to genotype-based antimicrobial resistance (AMR) prediction. Whole genome sequencing (WGS) has revolutionized our understanding of bacterial genomes, providing a comprehensive view of the genetic determinants of AMR \autocite{Mustafa2024-bs}. Early computational pipelines such as \textbf{ResFinder} \autocite{Florensa2022-jf}, \textbf{ARG-ANNOT} \autocite{Gupta2014-ue}, and \textbf{CARD} \autocite{Alcock2023-gh} provided curated resistance gene catalogs linked to antibiotic classes. These resources accelerated the in silico prediction by mapping known resistance determinants to genomic sequences using \textbf{sequence similarity} or \textbf{BLAST}-based approaches. However, such alignment-dependent methods struggled to generalize to novel or low-homology sequences and could not infer resistance mechanisms from previously un-characterized mutations.

To overcome these limitations and harness this wealth of genomic information the field increasingly adopted machine learning (ML) and deep learning (DL) models that leverage statistical patterns across genomic features. Early approaches employed feature engineering — such as \textbf{k-mer frequencies}, \textbf{SNP profiles}, or \textbf{gene presence-absence matrices} — as input to classifiers like Random Forests, Support Vector Machines (SVMs), and Gradient Boosting algorithms \autocite{Hu2024-qx}, \autocite{Valavarasu2025-js}. These machine learning algorithms, driven by the availability of larger datasets and improved computational capabilities, offer a promising opportunity to utilize the WGS data for predicting AMR with enhanced accuracy and efficacy \autocite{Liu2020-hi}. Moreover, ML-based in silico methods hold immense potential to transform clinical practice by facilitating rapid, scalable, and data-driven AMR diagnostics, enabling clinicians to tailor antibiotic therapies to the resistance profiles of infecting pathogens. Building on this foundation, recent work has employed deep learning strategies, such as transfer learning convolutional neural networks (CNNs) \autocite{Rayesha2025-aa}, \autocite{10957126}, to address challenges like data imbalance, where resistant strains are underrepresented due to antibiotic-specific variations, a challenge likely to persist as novel antimicrobials emerge.

The emergence of deep learning architectures marked a significant shift. DeepARG \autocite{Arango-Argoty2018-ej} used convolutional neural networks (CNNs) trained on metagenomic reads to classify antibiotic resistance genes (ARGs) with high precision, outperforming alignment-based methods on unseen data. Similarly, AMRPlusPlus and DeepAMR extended this paradigm by incorporating multi-task learning, where one model simultaneously predicts resistance to multiple antibiotic classes. Various algorithms such as \textit{Graphing Resistance Out Of meTagenomes (GROOT)} \autocite{Rowe2018-kn} and \textit{Meta-MARC} \autocite{Lakin2019-lp} further integrated resistome prediction with read-level assembly, enabling rapid resistome screening in metagenomic samples. Collectively, these advancements demonstrate that deep neural architectures can infer high-level genomic representations relevant to resistance phenotypes, establishing the foundation for next-generation AMR predictive frameworks.

Despite the inherent capabilities of ML-based AMR prediction strategies, most current models remain limited by their reliance on \textbf{single nucleotide polymorphisms (SNPs)} as the primary genomic feature set \autocite{Ren2022-ao}, \autocite{Ren2022-uf}. That is, while these data-driven approaches improved detection accuracy, they remained \textbf{static} — modeling resistance as a fixed phenotype derived from a snapshot of the genome. In reality, bacterial populations evolve dynamically under selective pressures, acquiring resistance through HGT events \autocite{Sevillya2020-iz}, plasmid acquisition, recombination, and point mutations \autocite{Von_Wintersdorff2016-dk}. HGT allows the bacteria to acquire resistance genes from other organisms, significantly accelerating the spread of AMR \autocite{Sun2019-us}. Consequently, the single-locus approaches may not adequately account for these critical factors influencing the evolution of AMR \autocite{Sevillya2020-iz}. Furthermore, the complex interactions between genetic mutations can contribute to emergent resistance phenotypes that are not captured by single-locus analysis. Conventional ML models trained on static datasets cannot capture this evolutionary flux or anticipate emergent resistance patterns that arise through novel genetic combinations. To address these limitations, there is growing recognition that predictive AMR modeling must account for evolutionary and ecological processes rather than solely genomic features \autocite{Bottery2021-px}.

\subsubsection{\textbf{Evolutionary Modeling and Simulation of Resistance}}
\label{subsubsec:evol_ml}
The evolutionary nature of antimicrobial resistance (AMR) has long been studied through the lens of population genetics and mathematical modeling. Foundational frameworks, such as the \textbf{Wright–Fisher model} and the \textbf{Moran model}, have been employed to describe the fixation probabilities of resistance alleles under antibiotic‐driven selection \autocite{alexandre2024bridgingwrightfishermoranmodels}. More mechanistic approaches, for instance those invoking the concept of fitness landscapes, have visualized how mutations alter growth rates or survival probabilities across varying drug concentrations \autocite{Das2020-zh}, \autocite{Harmand2017-vb}. These models have elucidated key principles of resistance evolution—such as the role of compensatory mutations in offsetting fitness costs and the stepwise acquisition of multi-drug resistance phenotypes.

Beyond analytical models, computational evolutionary simulations have gained traction for exploring resistance pathways. \textit{Agent-based models (ABMs)} simulate individual bacteria within structured environments, capturing stochastic mutation events, spatial dynamics, and gene transfer mechanisms \autocite{Chait2017-qy}. Similarly, genetic algorithms and evolutionary strategies have been applied to approximate natural selection by iteratively evolving populations according to fitness functions \autocite{Ragalo2018-sk}. 

In addition to using ML/DL frameworks, phylogenetic analysis and heuristic-based approaches have also been leveraged for AMR detection and prediction. Reference \autocite{Yurtseven2023-vf} employs a pipeline that integrates a \textbf{phylogeny-related parallelism} score to filter mutations linked to population structure. This pre-processing step precedes the training of support vector machines (SVMs) and random forests (RFs) on whole-genome sequences to identify AMR markers. Conversely, heuristic-based methods such as genetic algorithms, previously deployed for driver gene prediction in cancer \autocite{He2022-tm}, offer a complementary avenue. By incorporating mutation and crossover operations, they enable heuristic feature selection and model optimization, thereby capturing subtle genetic interactions and potentially mitigating biases inherent to phylogenetic inference.

Moreover, genetic algorithms have shown promise in various research domains, such as genotype and phenotype prediction. Reference \autocite{Mowlaei2023-uh} introduces \textbf{FSF-GA}, a feature selection framework tailored for phenotype prediction of quantitative traits. By reducing feature space and identifying relevant genotypes, this hybrid approach integrates pre-processing and genetic algorithm-based selection to predict optimal SNP combinations for trait prediction. By incorporating SNPs and leveraging genetic interactions, they can facilitate the predictive evolution of genes towards desired phenotypes, addressing the challenges posed by complex genetic interactions.

More recently, the dual challenge of tracking AMR spread through \textbf{phylogenetic reconstruction} and optimizing treatment strategies via computational techniques such as GAs has led to the development of newer frameworks \autocite{Colin2020-zx}, \autocite{Yurtseven2023-lf}. These modeling systems enable both the inference of evolutionary trajectories and the precise prediction of resistance phenotypes in clinically relevant pathogens.

HGT modeling has also been critical in simulating resistance dissemination across species boundaries. Studies using network-based approaches have shown that plasmid-mediated HGT accelerates the spread of resistance genes within microbial communities \autocite{Brito2021-tp}, \autocite{Wang2024-zm}. Large-scale studies of plasmid genomes have revealed how \textbf{antibiotic resistance genes (ARGs)} move across plasmids and host taxa, underscoring the importance of \textit{conjugative mobile genetic elements} in AMR propagation \autocite{Vrancianu2020-jq}. Parallel to this, simulation frameworks such as \textbf{SimBac} \autocite{Brown2016-ct} allow for whole-genome bacterial evolution modeling under homologous recombination and inter-species gene transfer, thereby reconstructing mutation and recombination histories across bacterial lineages \autocite{Didelot2007-tv}.

While these models offer valuable insights into evolutionary dynamics, most operate independently from data-driven machine learning pipelines. That is, they rely on manually defined fitness parameters or theoretical assumptions about selective pressures, rather than being learned directly from genomic data. This disconnect limits their predictive power and adaptability to real-world genomic datasets, where complex, non-linear dependencies govern resistance acquisition. Bridging this gap requires hybrid frameworks that \textbf{embed ML-derived fitness estimations} within evolutionary simulations, enabling data-informed evolutionary trajectories.

\subsubsection{\textbf{Hybrid Multi-Objective Computational Frameworks}}
Recent years have witnessed increasing efforts to integrate evolutionary computation and machine learning for biological inference. Evolutionary algorithms (EAs) have been widely adopted in bioinformatics for tasks such as feature selection, drug design, and protein-ligand docking. The Mixture of Experts (MoE) architecture, originally proposed for ensemble learning \autocite{Jacobs1991-ir}, provides a mechanism to combine multiple specialized models (\textbf{``experts"}) that focus on distinct data subspaces, with a gating network determining their relative contributions. MoE architectures have recently gained prominence in bioinformatics for modeling heterogeneous biological data. They have been effectively applied to multi-omics data integration \autocite{Minoura_etal_2021_scMM}, \autocite{Spooner2025-tt}, enabling the fusion of transcriptomic, proteomic, and metabolomic modalities; to gene-expression analysis, where expert subnetworks capture condition-specific regulatory patterns; and to protein-sequence classification, leveraging modular expert specialization for learning discriminative sequence representations \autocite{Sun_etal_2024_AIDOProtein}, \autocite{Guan2022-js}.

Despite their individual successes, very few studies have sought to couple MoE frameworks with evolutionary algorithms in the context of AMR or microbial evolution. Some precedents exist in other domains, such as \textbf{Neuroevolution of Augmenting Topologies (NEAT)} \autocite{Stanley2002-lk} and \textbf{CoDeepNEAT} \autocite{miikkulainen2017evolvingdeepneuralnetworks} demonstrated the synergy between gradient-based and evolutionary optimization for neural architecture search. Similarly, \textbf{Evolutionary Reinforcement Learning (ERL)} and \textbf{Genetic Programming (GP)} have shown that hybridizing ML with evolutionary dynamics can uncover robust solutions in high-dimensional, noisy fitness landscapes \autocite{khadka2018evolutionguidedpolicygradientreinforcement}. These successes suggest that combining data-driven learning with biologically inspired evolution could yield predictive frameworks for resistance emergence.

In AMR research, a select number of recent studies have begun exploring integrative paradigms that combine evolutionary inference with machine-learning approaches. For example, one modeling effort projected the development of resistance in \textit{Neisseria gonorrhoeae} using mathematical and surveillance-informed phylogenetic techniques \autocite{Riou2023-uw}. In another study, a deep reinforcement learning framework called \textbf{AMPainter} was used to generate antimicrobial peptides by evolving sequences in silico, although the evolution was treated purely as a generative process rather than as a mechanistic simulation of microbial population dynamics \autocite{Dong2025-rt}. Despite these advances, such methods still typically consider evolution as an external constraint or generative prior, instead of embedding it as an active simulation of resistance development governed by a learned fitness landscape.

The \textbf{Evolutionary Mixture of Experts (EvoMoE)} proposed in this study extends these ideas by embedding a machine learning–based AMR prediction model directly within a genetic algorithm, such that the learned resistance probability functions as the fitness measure for simulated evolution. This allows the system to not only classify resistance states but also dynamically explore hypothetical evolutionary trajectories leading to resistance acquisition. By tracking the distribution of predicted probabilities across generations, EvoMoE can identify evolutionary attractors, genomic configurations toward which resistant populations converge, offering new insights into the mechanistic underpinnings of AMR development.

\subsubsection{\textbf{Gap Analysis and Research Motivation}}
\label{subsubsec:gap_motive}
From the synthesis of prior literature, several key limitations emerge:
\begin{itemize}
    \item \textbf{Static Genomic Predictors}: Most ML-based AMR models treat genomes as static entities, lacking mechanisms to account for temporal or evolutionary changes.
    \item \textbf{Simplified Evolutionary Models}: Existing evolutionary frameworks use hand-crafted or fixed fitness functions, without leveraging data-driven AMR predictions to shape selection dynamics.
    \item \textbf{Absence of Hybridization}: There is minimal exploration of integrating ML (for prediction) and GA (for simulation) within a unified, feedback-driven system that emulates biological evolution.
    \item \textbf{Limited Interpretability of Evolutionary Pathways}: Few studies visualize or quantify how genomes evolve toward resistance, limiting their use in hypothesis generation for drug discovery.
\end{itemize}
The EvoMoE framework introduced in this paper directly addresses these gaps by merging supervised genomic learning with stochastic evolutionary optimization. The MoE model serves as a high-fidelity estimator of AMR probability, which in turn drives the genetic algorithm’s selection process. This creates a feedback loop where learning guides evolution — and evolution reveals patterns that inform our understanding of resistance emergence. Moreover, sensitivity analyses on parameters such as mutation rate and selection pressure allow for systematic exploration of how evolutionary conditions modulate resistance pathways, providing both mechanistic insight and translational value.

\subsection{Objective}
\label{subsec:objective}
The goal is to develop an in silico evolutionary system that integrates machine learning and genetic algorithms within a unified, feedback-driven loop. In this framework, the GA simulates microbial evolution under antibiotic selection, while an \textbf{EvoMoE}-based fitness function, trained to predict AMR potential from genomic sequences, guides the evolutionary search toward resistant phenotypes. Specifically, the proposed approach, centered on \textit{Escherichia coli} as a model organism, aims to move beyond static genome-based prediction toward a mechanistic understanding of resistance evolution.

Additionally, the framework incorporates a probabilistic crossover mechanism to emulate HGT, thereby capturing the genetic exchanges that accelerate resistance dissemination across microbial populations. By embedding this stochasticity into the evolutionary model, the system enhances diversity and realism in simulated evolutionary pathways. The proposed framework is designed to achieve two main outcomes:
\begin{enumerate}
    \item \textbf{Prediction of resistant genotypes} — accurately identifying genomic configurations associated with resistance to specific antibiotics.
    \item \textbf{Reconstruction of evolutionary trajectories} — mapping the mutational and transfer events leading to resistance acquisition, alongside visualizing the corresponding changes in the fitness landscape.
\end{enumerate}

\vfill
\pagebreak

\section{Methods}
\label{sec:methods}
We leverage the \textbf{MicroBIGG-E database} \autocite{Feldgarden2022-pi}, a curated and comprehensive repository of microbial genomic data, as the foundational dataset for our study. To ensure species-level consistency and streamline evolutionary analysis, we restrict our investigation to \textit{Escherichia coli} isolates. A collection of \textit{E. coli} gene sequences serves as the initial population for our evolutionary simulations.

A Genetic Algorithm (GA) is used to emulate the evolutionary trajectories of these genes, allowing for the progressive accumulation of point mutations and horizontal gene transfer (HGT)–like events. Through iterative selection, crossover, and mutation, the GA explores the fitness landscape underlying potential \textbf{resistance-conferring genotypes}. This process thereby provides a computational model of microbial evolution under selective antibiotic pressure.

\begin{figure}[b]
\begin{center}
\includegraphics[width=8cm]{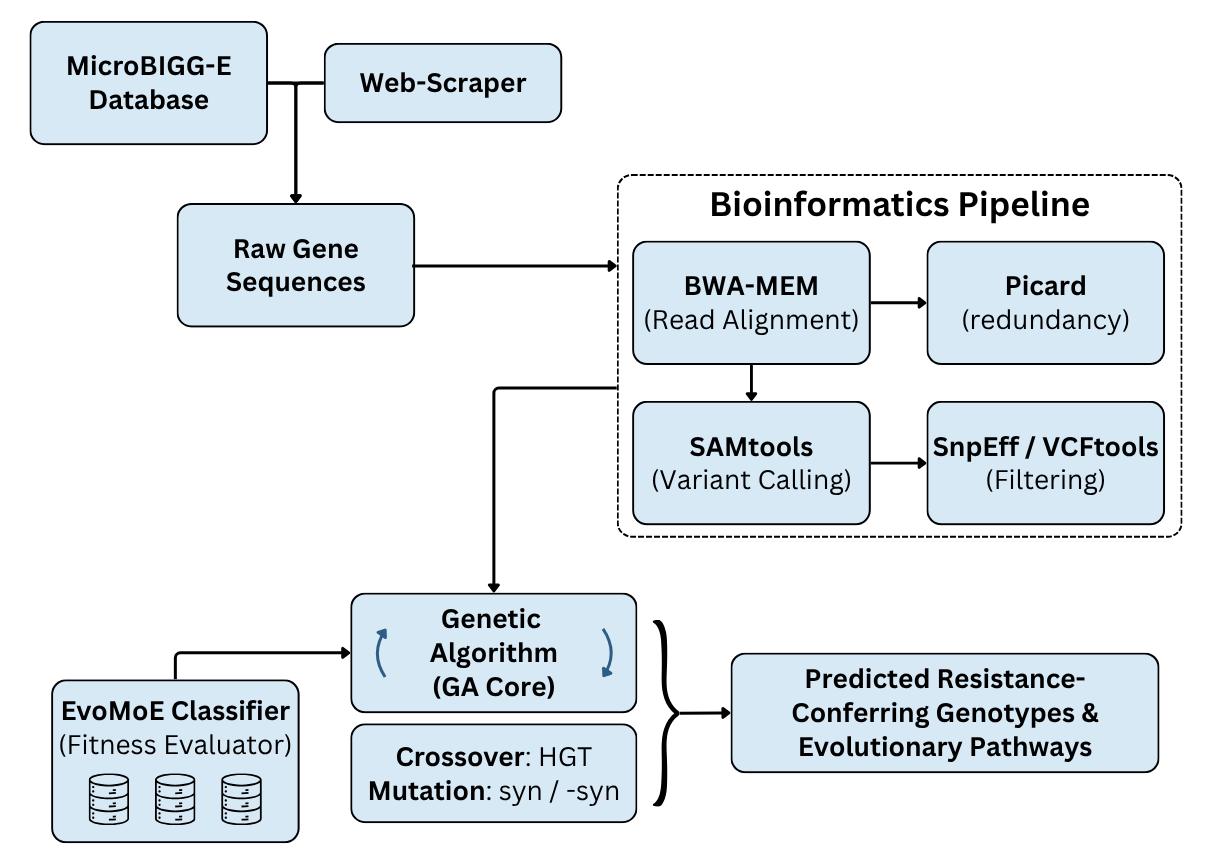}   
\caption{\textbf{Brief Overview of the AMR-MoEGA Computational Framework.}\\ The workflow integrates a Genetic Algorithm (GA) for evolutionary search with a Machine Learning (ML)-driven fitness evaluation. Sequence data from MicroBIGG-E and a Web Scraper are processed into feature strings. Bioinformatics tools (BWA-mem, SAMtools, VCFtools) generate the SNP feature matrices used to train the MoE classifier, which serves as the fitness function for the GA's exploration of resistance-conferring genotypes.}
\label{fig:flowchart}
\end{center}
\end{figure}

A schematic overview of the proposed workflow is seen in \hyperref[fig:flowchart]{Figure \ref{fig:flowchart}}. A central innovation of this framework lies in the design of the fitness function, which governs the GA’s evaluation of candidate genomes. Traditional approaches often rely exclusively on static SNP-based metrics to approximate resistance potential. In contrast, we integrate a machine-learning-driven fitness evaluation by coupling the GA with an \textbf{MoE classifier} trained on sequence-derived SNP matrices. Prior to training, raw gene sequences are processed through a bioinformatics pipeline employing standard tools for read alignment (e.g., BWA), variant calling (e.g., SAMtools), and functional annotation (e.g., SnpEff). This pipeline yields high-dimensional SNP feature matrices that capture both coding and non-coding variation relevant to antimicrobial resistance. The trained XGBoost model predicts AMR phenotypes for individual genotypes, and these predictions directly inform the GA’s fitness assessment, effectively linking evolutionary search with phenotypic inference.

To enhance biological realism, probabilistic crossover operations are designed to simulate HGT events, while mutation functions capture both synonymous and non-synonymous substitutions. This combination allows the algorithm to explore a broader and more biologically plausible genomic landscape, facilitating insight into evolutionary pathways that may underlie emerging resistance mechanisms. \footnote{https://github.com/anshul-2010/AMR-MoEGA}

Together, these components constitute AMR-MoEGA, a hybrid Mixture of Experts + Genetic Algorithm framework, that unites evolutionary computation with data-driven phenotype prediction to model the genomic evolution of antimicrobial resistance. The remainder of this section is structured as follows:
\begin{itemize}
    \item \hyperref[subsec:data]{Section \ref{subsec:data}} describes data acquisition, filtration, and preprocessing steps ensuring genomic data quality.
    \item \hyperref[subsec:bioinfo]{Section \ref{subsec:bioinfo}} details the bioinformatics workflow for SNP extraction and feature generation.
    \item \hyperref[subsec:xgboost]{Section \ref{subsec:xgboost}} outlines the supervised learning approach used for AMR phenotype prediction.
    \item \hyperref[subsec:ga]{Section \ref{subsec:ga}} explains the implementation of the genetic algorithm, including evolutionary operators, fitness evaluation, and convergence criteria.
\end{itemize}

\subsection{Data Acquisition and Filtration}
\label{subsec:data}
The MicroBIGG-E database \autocite{Feldgarden2022-pi} served as the foundational resource for our in silico exploration of antimicrobial resistance (AMR) evolution in \textit{Escherichia coli}. MicroBIGG-E provides an extensive, curated repository of microbial genomic data, integrating annotations of AMR genes, virulence factors, and associated metadata derived from \textbf{GenBank} submissions. This database enables systematic identification and comparative analysis of genetic determinants implicated in AMR across diverse bacterial populations.

To construct a focused and biologically consistent dataset, we implemented a \textbf{multi-tiered filtration strategy} within the MicroBIGG-E platform. Each filtration layer, as shown in \hyperref[Table_microbigge]{Table \ref{Table_microbigge}}, was designed to progressively refine the dataset toward genomic regions most relevant to resistance mechanisms and human-pathogenic \textit{E. coli} lineages.

\subsubsection{\textbf{Type Filter — AMR}} The primary selection criterion targeted entries classified as Type: AMR, thereby isolating genes explicitly annotated as antimicrobial resistance determinants. This ensured that subsequent analyses were restricted to sequences functionally relevant to resistance phenotypes, excluding unrelated virulence or housekeeping genes.
\subsubsection{\textbf{Method Filter — BlastP}} To enhance annotation accuracy, we applied the Method: BlastP filter. BlastP performs sequence alignment at the protein level, identifying homologs with significant similarity to experimentally validated AMR proteins. Restricting to BlastP-derived annotations increased confidence in functional homology and minimized inclusion of spurious or weakly characterized genes.
\subsubsection{\textbf{Strand Filter — Positive}} We further refined the dataset using the Strand: Positive filter, restricting selection to genes encoded on the forward ($5'\rightarrow3'$) strand. While both strands can encode functional genes, this choice standardized orientation across samples, simplifying downstream feature encoding and reducing directional variance in sequence-based analysis.
\subsubsection{\textbf{Host Filter — Homo sapiens}} To ensure clinical relevance, we employed the Host: Homo sapiens criterion, selecting only E. coli isolates associated with human infections. This constraint aligns the dataset with AMR patterns of medical significance, emphasizing genotypes with demonstrated or potential pathogenicity in human hosts.
\subsubsection{\textbf{Scope Filter — Core Genome}} Finally, we applied the Scope: Core filter to isolate genes comprising the core genome of E. coli—that is, genes conserved across nearly all strains and essential for basic cellular function. Focusing on core genes reduces confounding effects from strain-specific accessory elements and facilitates the study of evolutionary modifications within conserved, biologically fundamental loci.

\begin{table}[t]
\caption{MicroBIGG-E Data Filtration}
\begin{center}
\begin{tabular}{|c|c|c|c|}
\hline
{\textbf{Data Parameters}} & {\textbf{Data filters chosen}} \\
\hline
Type & Antimicrobial Resistance (AMR) \\
\hline
Method & BlastP\\
\hline
Strand & Positive (+)\\
\hline
Host organism & \textit{Homo sapiens}\\
\hline
Scope & Core\\
\hline
Organism & \textit{Escherichia Coli}\\
\hline
\end{tabular}
\label{Table_microbigge}
\end{center}
\end{table}
Following this filtration pipeline, we obtained a curated pool of \textit{E. coli} AMR-associated genes representing conserved resistance-linked genomic regions across human-pathogenic isolates. From this refined dataset, a random subset of 100 genes was selected to initialize each genetic algorithm (GA) simulation. This sampling strategy balances computational efficiency with genetic diversity, ensuring sufficient representation of mutational and recombinational dynamics during the evolutionary modeling phase.

\subsection{Bioinformatics Tools}
\label{subsec:bioinfo}
\begin{figure}[b]
\begin{center}
\includegraphics[width=9cm]{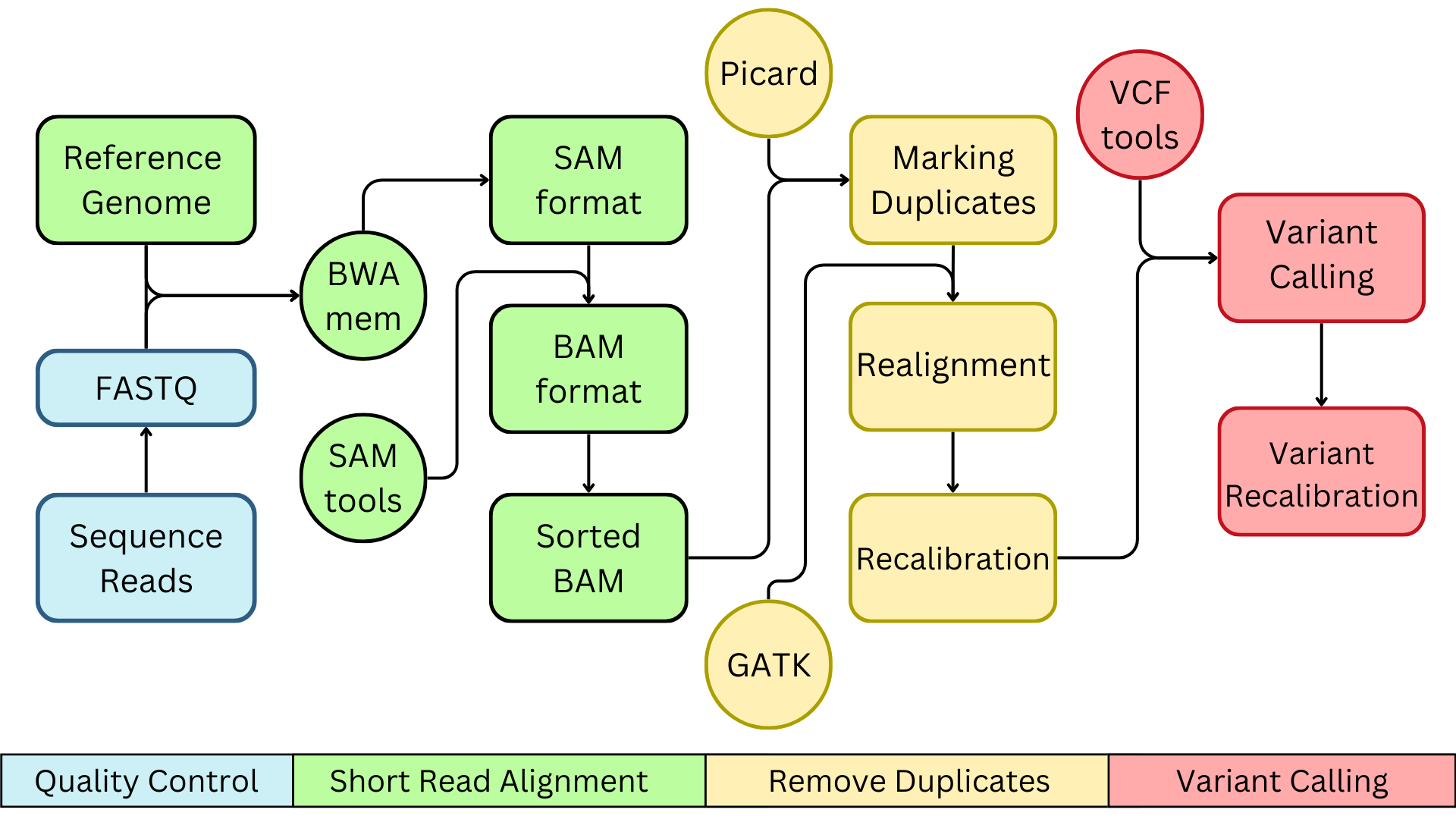}   
\caption{\textbf{Bioinformatics Pipeline for Reads Processing and Variant Calling.}\\ The workflow details the sequential steps from raw sequence reads (FASTQ) to final recalibrated variants. Reads are aligned to a Reference Genome using BWA-MEM, converted to BAM format and sorted by SAMtools. Quality control and read cleanup, including marking duplicates and realignment (Picard and GATK), precede the final Variant Calling and Recalibration steps to ensure high-confidence SNP and indel identification.}
\label{fig:bioinformatics}
\end{center}
\end{figure}

In our preprocessing pipeline, accurate interpretation of information embedded within the genetic sequences is paramount. To extract biologically meaningful insights and prepare the data for downstream modeling, we implemented a comprehensive bioinformatics workflow integrating read-level quality control, genomic alignment, variant discovery, functional annotation, and feature matrix construction.  A brief pipeline for this workflow is shown in \hyperref[fig:bioinformatics]{Figure \ref{fig:bioinformatics}}. By leveraging a suite of validated bioinformatics tools at each stage, this pipeline ensures both analytical rigor and compatibility with the subsequent machine learning based fitness evaluation within the AMR-MoEGA framework.

\subsubsection{\textbf{Quality Assessment and Read Trimming}} Raw sequencing reads were subjected to quality control (QC) using \textbf{FastQC} \autocite{andrews2010fastqc}, which evaluates key metrics including per-base sequence quality, GC content, adapter contamination, and the presence of over-represented k-mers. Based on these QC reports, reads were processed using \textbf{Trimmomatic} to remove low-quality bases (Phred score $<$ Q15) and residual adapter sequences from the $3'$ and $5'$ ends. This step mitigates potential downstream biases arising from sequencing artifacts, ensuring that only high-confidence reads proceed to alignment.

\subsubsection{\textbf{Sequence Alignment and Preprocessing}} Following quality filtration, the high-confidence reads were aligned to the \textit{Escherichia coli K-12 MG1655} reference genome using the \textbf{BWA-MEM algorithm} \autocite{li2013aligning}. Prior to alignment, the reference genome was indexed using BWA’s built-in indexing module. Default parameters were used, allowing for up to four mismatches per read while accommodating small indels \autocite{Li2009-zq}.
The resulting SAM alignment files were converted into BAM format, sorted by genomic coordinates, and indexed using \textbf{SAMtools} \autocite{Li2009-ou}. Duplicate reads, arising from PCR amplification, were subsequently marked using \textbf{Picard} \autocite{Zhao2018-yi}. These preprocessing steps reduced redundancy and optimized data structures for efficient variant detection.

\subsubsection{\textbf{Variant Calling and Refinement}} Variant discovery was performed using the \textbf{Genome Analysis Toolkit (GATK)} \autocite{Van_der_Auwera2013-uq}. The pipeline included local realignment around indels and base quality score recalibration (BQSR) to correct systematic biases in quality scoring. Variants were called in SNP and indel discovery mode, producing high-resolution \textbf{Variant Call Format (VCF)} files.
To ensure accuracy, we applied variant-level filtering using \textbf{VCFtools} \autocite{Danecek2011-aw}, enforcing thresholds on minimum read depth ($\geq$ 10×), mapping quality (MQ $\geq$ 30), and genotype quality (GQ $\geq$ 50). These filters eliminated low-confidence variants arising from ambiguous alignments or sequencing noise.

\subsubsection{\textbf{Functional Annotation of Variants}} Filtered variants were functionally annotated using \textbf{SnpEff}, which predicts the biological effects of nucleotide substitutions on coding and regulatory regions. Each SNP was classified as synonymous, non-synonymous, nonsense, or missense, with additional annotations linking variants to gene ontology (GO) terms and known AMR-related functional pathways where available. This annotation process contextualized the detected genomic variation, facilitating interpretation of potential resistance-associated mutations.

\subsubsection{\textbf{SNP Feature Matrix Construction}} The annotated variants were systematically encoded into SNP feature matrices, where each row represented an individual gene instance and each column corresponded to a distinct variant locus or feature. Binary encoding captured variant presence/absence, while continuous features such as allele frequency and quality scores were retained as quantitative attributes. To maintain comparability across genes and isolates, all feature matrices were z-score normalized prior to downstream analysis.
This structured representation captured both genetic diversity and functional relevance, forming the basis for subsequent machine learning–driven phenotype prediction.

\subsection{Mixture of Experts framework for AMR Classification}
\label{subsec:xgboost}
To address the heterogeneity and complex feature dependencies inherent in antimicrobial resistance (AMR) genomic data, we designed a \textbf{Mixture of Experts (MoE)} ensemble framework that integrates multiple complementary learning paradigms. The rationale behind this approach stems from the observation that distinct classifiers often excel in modeling different structural aspects of genomic variation—nonlinearity, feature sparsity, and hierarchical dependencies. By jointly leveraging their strengths, the MoE framework aims to achieve a robust and generalizable AMR classification model.

\subsubsection{\textbf{Feature Matrix Construction}} Following the bioinformatics preprocessing steps outlined in \hyperref[subsec:bioinfo]{Section \ref{subsec:bioinfo}}, reference alleles, variant alleles, and their corresponding positions are extracted from the Giessen dataset. The isolates are then merged based on reference allele position, creating a consensus view of all genetic variation at each position. The loci devoid of variation (represented by \textbf{N}) are excluded. The final output of the above steps is a SNP matrix, where the rows represent individual isolates and columns represent variant alleles. Label encoding is employed to transform the categorical nucleotide symbols (\textbf{A}, \textbf{C}, \textbf{G}, \textbf{T}) and \textbf{N} symbol for missing variants into numerical features suitable for the machine learning model (A = 1, G = 2, C = 3, T = 4, N = 0). 

\subsubsection{\textbf{LLM-based Genomic Embedding Integration}} 
While the SNP-derived feature matrix offers a discrete and interpretable representation of genomic variation, it may not fully capture the higher-order dependencies and contextual semantics present within genomic sequences. To enrich the feature space for the classification component, we incorporated \textbf{pretrained large language model (LLM)} embeddings derived from biological sequence model \textbf{DNABERT}. This model encode genomic subsequences into dense contextual vectors by learning co-occurrence and positional dependencies among k-mer tokens.

For each isolate, relevant AMR-associated genomic regions were tokenized into overlapping k-mers and passed through the pretrained LLM to obtain embedding vectors. These vectors were concatenated with the SNP-based meta-features to form a hybrid representation that retains explicit variant-level information while introducing sequence-level contextual awareness. The resulting hybrid embeddings were used exclusively for the \textbf{Mixture of Experts (MoE) classification framework}, enhancing its ability to discern subtle resistance-associated sequence patterns.

It is important to note that downstream modules, particularly the \textbf{Genetic Algorithm (GA)}-based evolutionary modeling, continued to operate on the canonical SNP-encoded feature space. This separation ensures compatibility with genetic operators such as mutation and crossover, which are inherently defined on discrete representations rather than continuous embeddings.

\subsubsection{\textbf{Expert Models}} 
\begin{figure*}[t]
  \includegraphics[width=\textwidth]{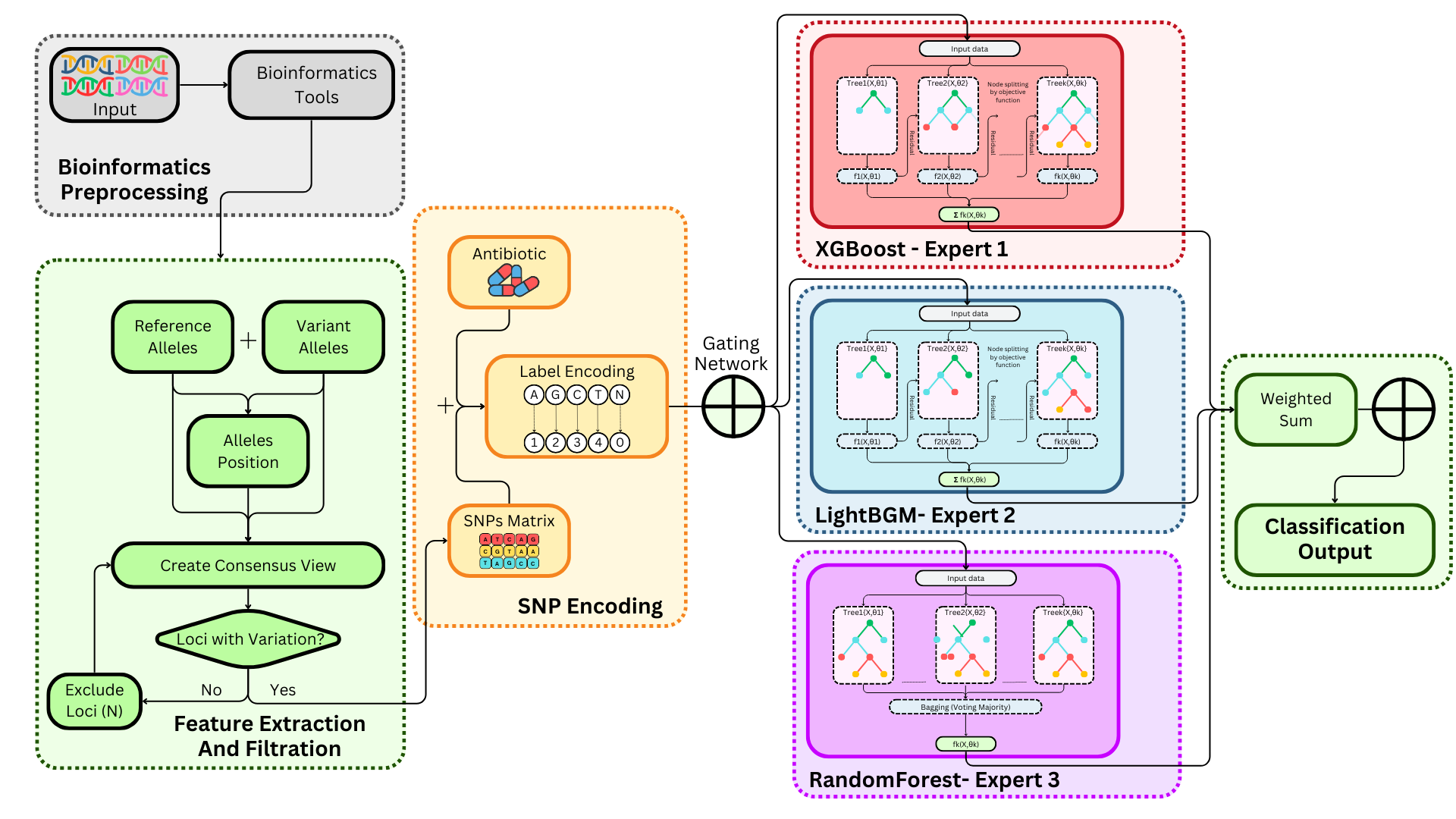}
  \caption{\textbf{Schematic overview of the proposed Mixture of Experts (MoE) framework for AMR classification.}\\
  Genomic variants from the MicroBIGG-E database undergo preprocessing, SNP matrix construction, and feature encoding. Three expert models: XGBoost, LightGBM, and Random Forest, independently learn distinct patterns from the genomic feature space. An adaptive gating network assigns instance-specific weights to each expert’s output, and the final AMR prediction is obtained through a weighted fusion of expert responses. This design enhances predictive robustness, interpretability, and generalization across E. coli isolates.}
  \label{fig:moe_framework}
\end{figure*}

The MoE framework as seen in \hyperref[fig:moe_framework]{Figure \ref{fig:moe_framework}} is composed of three base experts: \textbf{XGBoost}, \textbf{LightGBM}, and \textbf{Random Forest}, chosen for their proven capacity to model nonlinear interactions and heterogeneous feature importance distributions.
\begin{itemize}
    \item \textbf{XGBoost} serves as the gradient boosting expert that efficiently captures additive non-linearities in the data.
    \item \textbf{LightGBM} complements this by employing its Gradient-based One-Side Sampling (\textbf{GOSS}) and Exclusive Feature Bundling (\textbf{EFB}) strategies, which are well-suited for large and sparse SNP matrices.
    \item \textbf{Random Forest}, a bagging-based ensemble, contributes diversity and stabilizes the ensemble against overfitting on dominant feature clusters.
\end{itemize}
Each expert is trained independently using optimized hyperparameters derived through \textbf{Bayesian optimization with cross-validation}, ensuring model-specific convergence and balance between bias and variance.
\begin{figure}[b]
\begin{center}
\includegraphics[width=7cm]{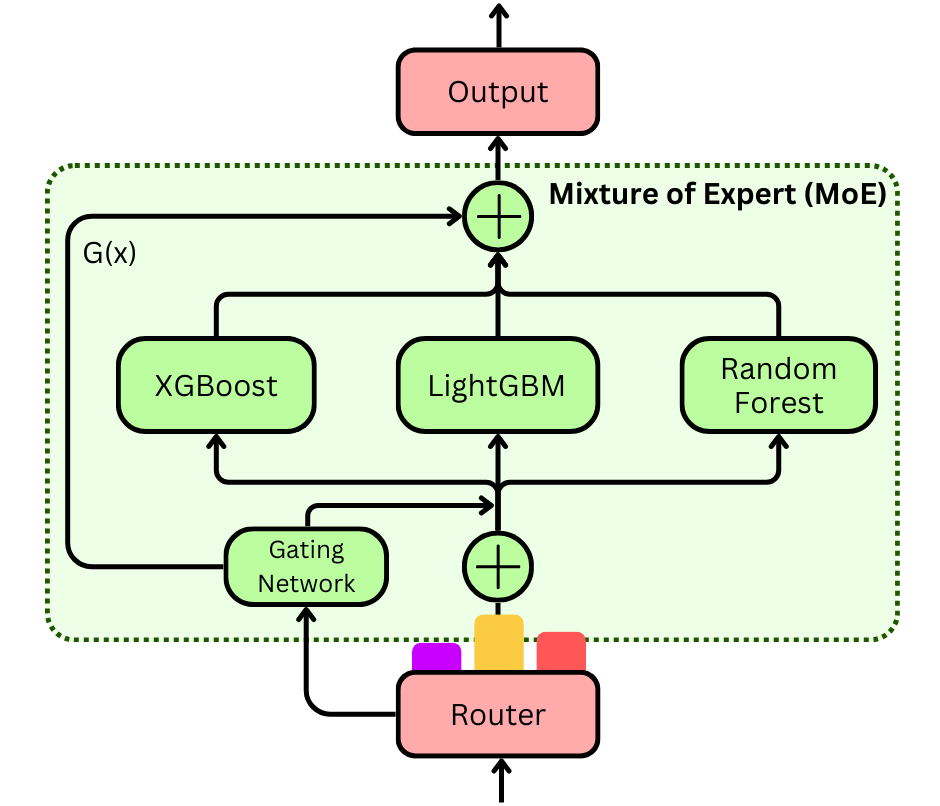}   
\caption{\textbf{Architecture of the Adaptive Layer of MoE.}\\ The model is comprised of a Router that distributes input data to a suite of specialized Expert models (XGBoost, LightGBM, Random Forest). A key innovation is the Gating Network, which adaptively learns to weigh the predictions of each expert based on the input features, $G(x)$, before summing them to produce the final classification Output.}
\label{fig:gating_network}
\end{center}
\end{figure}

\subsubsection{\textbf{Adaptive Gating Network}} Unlike conventional averaging or static weighting schemes, our approach introduces an \textbf{adaptive gating network}, implemented as a lightweight feedforward neural network that dynamically learns to weight expert predictions based on the input feature representation. 

The \hyperref[fig:gating_network]{Figure \ref{fig:gating_network}} shows how the gating mechanism has been incorporated in the workflow. Formally, for an input (x), the gating network computes a set of weights ($w_i(x)$), one per expert, such that:
\begin{equation} \label{eq:conditions}
\sum_{i=1}^{n} w_i(x) = 1, \quad w_i(x) \geq 0
\end{equation}
and the final prediction is:
\begin{equation} \label{eq:prediction_final}
\hat{y}(x) = \sum_{i=1}^{n} w_i(x) f_i(x)
\end{equation}
where ($f_i(x)$) denotes the prediction of the ($i^{th}$) expert. 

This mechanism allows the framework to adaptively emphasize experts that perform better on specific genomic substructures; for example, one expert may specialize in modeling multi-drug resistance patterns, while another captures rare single-gene mutations.

To further enhance robustness, we integrate \textbf{uncertainty-aware gating}, where the gating network is regularized with the predictive entropy of expert outputs. This enables the model to down-weight uncertain predictions, improving reliability in borderline or low-confidence cases.

\subsubsection{\textbf{Regularization and Knowledge Fusion}} To enhance generalization and encourage the emergence of complementary expertise among individual learners, the Mixture-of-Experts (MoE) framework incorporates a multi-objective regularization strategy. Specifically, two biologically and statistically grounded regularization components: \textbf{consistency regularization} and \textbf{uncertainty regularization} are introduced into the training setup. The motive behind inclusion of these regularization terms is to stabilize training and promote meaningful knowledge sharing without collapsing expert diversity.

\subsubsection*{\textbf{Consistency Regularization}} To avoid overfitting and encourage shared representation learning, we impose a consistency constraint across expert embeddings. This regularization minimizes the divergence between the intermediate feature representations produced by different experts for the same input sample. Formally, it is expressed as the cumulative \textbf{Kullback–Leibler (KL)} divergence between all pairs of expert embeddings, ensuring that experts maintain partially aligned internal representations while still preserving their specialization in distinct genomic subspaces. Such consistency acts as a soft knowledge-sharing mechanism, fostering a degree of representational overlap that enables robust ensemble reasoning. In the below \hyperref[eq:consistency]{Equation \ref{eq:consistency}}, $\sigma$ denotes the softmax function.
\begin{equation} \label{eq:consistency}
\mathcal{L}_{\text{consistency}} = \frac{1}{N(N-1)} \sum_{i \neq j} \text{KL}\big( \sigma(h_i(x)) || \sigma(h_j(x)) \big)
\end{equation}

\subsubsection*{\textbf{Uncertainty Regularization}} To regulate the gating mechanism and discourage over-reliance on unreliable experts, an uncertainty penalty is introduced. The uncertainty regularization term penalizes experts that exhibit \textbf{high predictive entropy}, encouraging the gating network to assign greater weight to confident and stable predictors. In practice, uncertainty can be quantified as the expected entropy of each expert’s probabilistic output or as the average KL divergence between an expert’s predictive distribution and the aggregated ensemble distribution. This formulation encourages experts to remain well-calibrated, reduces stochasticity in the gating process, and enhances the overall interpretability and stability of the ensemble. The influence of this term is modulated by the hyperparameter $\beta$, which balances the trade-off between predictive diversity and confidence-based expert selection.
\begin{align} \label{eq:uncertainty}
\mathcal{L}_{\text{uncertainty}} &= \sum_{i=1}^{N} w_i(x) H(f_i(x)) \\
&= -\sum_{i=1}^{N} w_i(x) \sum_{k} p_{i,k}(x) \log p_{i,k}(x)
\end{align}

The overall objective of the MoE framework combines individual expert losses with these two regularization components:
\begin{equation} \label{eq:overall_loss}
\mathcal{L}_{\text{total}} = \sum_i w_i(x) \mathcal{L}_i + \beta \mathcal{L}_{\text{uncertainty}} + \gamma \mathcal{L}_{\text{consistency}}
\end{equation}

where $w_i(x)$ denotes the gating weights for each expert, and $\beta$ and $\gamma$ are tunable coefficients controlling the contribution of the uncertainty and consistency regularization terms, respectively. This composite objective encourages experts to specialize meaningfully while remaining aligned under a coherent ensemble representation.

\subsubsection{\textbf{Model Integration and Evaluation}} The ensemble is trained in a two-stage pipeline designed to ensure both expert specialization and effective knowledge fusion.

\subsubsection*{\textbf{Expert Pretraining}} Each expert model is trained independently on the training subset of the genomic dataset. This allows each of the individual learners to capture distinct aspects of antimicrobial resistance (AMR) signatures, such as \textit{sequence motifs}, \textit{mutational dependencies}, or \textit{structural patterns}, without interference from the gating network.

\subsubsection*{\textbf{Gating Fine-tuning}} Once experts are pretrained, the gating network is trained using a held-out validation set to learn optimal mixture weights that dynamically combine expert outputs. The gating network leverages both expert predictions and intermediate representations to infer which expert is most informative for a given input, thereby achieving adaptive routing of genomic samples through the ensemble.

During inference, the MoE model outputs the predicted resistance score for each microbial isolate by aggregating expert predictions according to the learned gating weights. Model interpretability is achieved through two complementary mechanisms:
\begin{enumerate}[label=(\roman*)]
    \item \textbf{Feature importance aggregation} across experts, which reveals variant-level or gene-level drivers of resistance, and
    \item \textbf{Attention-weight visualization} from the gating network, which highlights expert-level contributions and contextual dependencies.
\end{enumerate}
Together, this integrated training and evaluation framework enables the MoE to act as both a high-accuracy predictive model and a biologically interpretable system capable of revealing mechanistic insights into antimicrobial resistance evolution.

\subsection{Genetic Algorithm Framework}
\label{subsec:ga}
As seen in \hyperref[sec:introduction]{Section \ref{sec:introduction}}, understanding the evolutionary dynamics underlying antimicrobial resistance (AMR) emergence requires models that capture both \textbf{mutation-driven} and \textbf{horizontal gene transfer (HGT)-mediated} mechanisms. To this end, we design a biologically informed \textbf{Genetic Algorithm (GA)} that emulates bacterial evolution under antimicrobial selective pressure. The GA simulates a population of bacterial genomes evolving over successive generations through processes analogous to natural selection, mutation, and gene transfer.

Each candidate solution represents a distinct bacterial genome encoded as a \textbf{chromosome}, a fixed-length string composed of nucleotides (A, C, G, T). The fitness of each genome ($S_i$) is evaluated using the \textbf{Mixture of Experts (MoE)} framework (Section \ref{subsec:xgboost}), which predicts the resistance potential of $S_i$ to a target antibiotic. This integration of ML-driven fitness estimation allows the GA to couple genotypic variation with phenotypic outcomes, providing a realistic simulation of resistance evolution.

\subsubsection{\textbf{Selection Mechanism}} Parent genomes for reproduction are chosen via \textbf{tournament selection}, a method balancing exploration and selective pressure. In each round, a subset of $k=5$ sequences is randomly sampled, and the genome with the highest fitness within the subset is selected as a parent. The selection probability $P_{sel}(S_i)$ is proportional to the fitness of $S_i$ relative to the population’s mean fitness ($F_{avg}$), defined as:
\begin{equation} \label{eq:selection}
P_{sel}(S_i) = \frac{F(S_i)}{t \times F_{avg}}
\end{equation}
where $t$ represents a normalization constant. This ensures that genomes exhibiting higher predicted resistance have greater likelihood of reproduction, reflecting the evolutionary principle of survival of the fittest.

\subsubsection{\textbf{Mutation Operator}} Mutation introduces controlled genetic diversity and mirrors spontaneous molecular changes in bacterial genomes. Unlike uniform random mutation, our mutation operator incorporates \textbf{biologically grounded biases} to better emulate real genomic variability. In bacterial evolution, mutation rates are not uniformly distributed across the genome - certain loci, known as \textbf{mutation hotspots}, such as \textit{CpG islands}, exhibit substantially elevated rates of nucleotide substitution compared to other regions \autocite{Hanson2022-gu}, \autocite{Van_Belkum1998-so}. Moreover, intrinsic mutation biases, including transitions-to-transversions asymmetry and codon usage preference, further influence the direction and magnitude of genetic variability. 

To capture these effects, we implement a probabilistic mutation model wherein mutation likelihoods are modulated by local sequence context, GC composition, and known hotspot annotations. This enables the algorithm to simulate realistic evolutionary dynamics, balancing stochastic exploration with biological plausibility. By embedding domain-specific mutation propensities within the genetic algorithm, we enhance its capacity to uncover adaptive trajectories that align with empirically observed AMR evolution in \textit{E. coli}.
\begin{itemize}
    \item \textbf{GC-Content Bias}: GC-rich regions are more prone to replication errors and mutational events. For each position ($i$) in the chromosome, the mutation probability ($P_{mut}(i)$) is modulated by local GC content using a sliding window approach:
    \begin{equation} \label{eq:gc}
    P_{mut}(i)=
    \begin{cases}
    f_{GC}(i), & \text{if } GC_{cont} > GC_{thresh}\\
    low\_mut\_prob, & \text{otherwise}
    \end{cases}
    \end{equation}
    where $f_{GC}(i)$ defines the functional dependence of mutation probability on local GC content.    
    \item \textbf{Transition/Transversion Bias}: Empirical studies show that \textit{transitions} (purine $\leftrightarrow$ purine or pyrimidine $\leftrightarrow$ pyrimidine) occur more frequently than \textit{transversions} (purine $\leftrightarrow$ pyrimidine). We model this using bias factors ($B_{transition}$) and ($B_{transversion}$):
    \begin{equation} \label{eq:trans}
    P(subs\_type) =
    \begin{cases}
    B_{transition}, & \text{if transition}\\
    B_{transversion}, & \text{if transversion}
    \end{cases}
    \end{equation}
    The overall per-site mutation probability is then:
    \begin{equation} \label{eq:mutation}
    P_{mutation}(i) = P_{mut}(i) \times P(subs\_type)
    \end{equation}
\end{itemize}
This composite model accounts for both local compositional effects and molecular substitution biases, yielding biologically plausible mutation dynamics.

\subsubsection{\textbf{HGT-Based Crossover and Synteny-Aware Exchange}} Horizontal Gene Transfer (HGT) is a dominant mechanism in microbial evolution, facilitating the exchange of genetic material—including antibiotic resistance genes (ARGs)—across species boundaries. Inspired by this biological phenomenon, we implement the synteny-guided, HGT-based crossover operator to replace or augment conventional recombination in the genetic algorithm (GA). This operator, based on a modified version of a previously established method \autocite{Sevillya2020-iz}, integrates genomic alignment and probabilistic gene transfer modeling to generate more biologically informed offspring. \hyperref[fig:hgt]{Figure \ref{fig:hgt}} schematically depicts the operator’s main stages—alignment, synteny evaluation, probabilistic transfer modeling, and offspring formation.

\begin{itemize}
    \item \textbf{Synteny Analysis and Neighborhood Conservation}: Each genome ($G = g_1, g_2, \ldots, g_n$) is modeled as an ordered sequence of genes, where each gene $g_i$ is associated with its nucleotide sequence. The \textit{k-neighborhood} of a focal gene ($g_0$) in G, denoted $N_k(G, g_0)$, comprises all genes within distance $k$ upstream or downstream of $g_0$. The \textbf{\textit{k-synteny index (k-SI)}} between two genomes ($G_i$ and $G_j$) quantifies the overlap in neighborhood composition, capturing local structural conservation of gene order.
    \begin{equation} \label{eq:SI_1}
        SI(g_0, G_i, G_j) = |N_k(G_i, g_0) \cap N_k(G_j, g_0)|
    \end{equation}
    High SI values indicate conserved gene neighborhoods and thus a lower likelihood of HGT, whereas low SI values (loss of synteny) increase the probability of transfer.
     
    \item \textbf{Global–Local Alignment and Segment Homology}: To identify candidate transfer segments, a contiguous region $S \subset G_i$ from a donor genome (parent $P_1$) is aligned against the recipient genome (parent $P_2$) using a global alignment scheme such as the \textbf{Needleman–Wunsch algorithm} \autocite{Needleman1970-yy}. This produces a homology score ($S_{g-l}(P_1^{seg}, P2)$). This score measures sequence-level compatibility across all possible insertion sites, where $l$ is the length of fixed segment that is a transfer candidate in the HGT simulation.
    
    \item \textbf{Synteny-Weighted Probabilistic Crossover}: Within each aligned region, we compute the local k-synteny score reflecting contextual similarity:
    \begin{equation} \label{eq:hgt}
    S_{k-syn}(G, g', k) = \sum_{g' \in N_k(G,g)} I(g', G)
    \end{equation}
    \begin{figure}[t]
        \begin{center}
        \includegraphics[width=8.4cm]{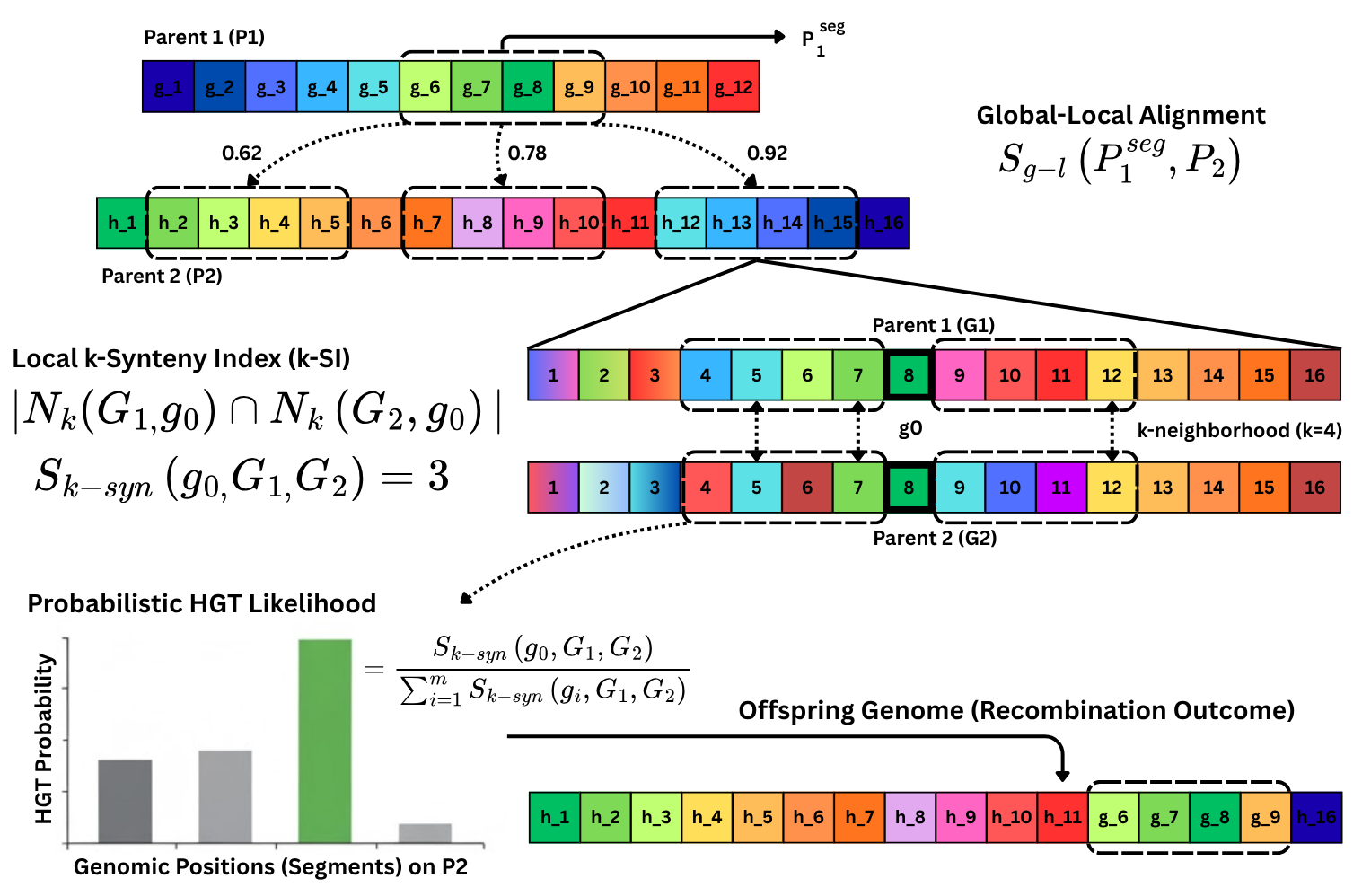}   
        \caption{\textbf{Synteny-guided HGT-based crossover.}\\
        A donor segment ($G_1$) from $P_1$ is aligned to segments from $P_2$ ($G_2$) and evaluated using both sequence similarity ($S_{g–l}$) and local gene order conservation ($S_{k–syn}$). \\
        The resulting probability model selects high-synteny regions for recombination, producing offspring genomes that mimic realistic HGT events.}
        \label{fig:hgt}
        \end{center}
    \end{figure}
    \begin{figure*}[t]
      \centering
      \includegraphics[width=\textwidth]{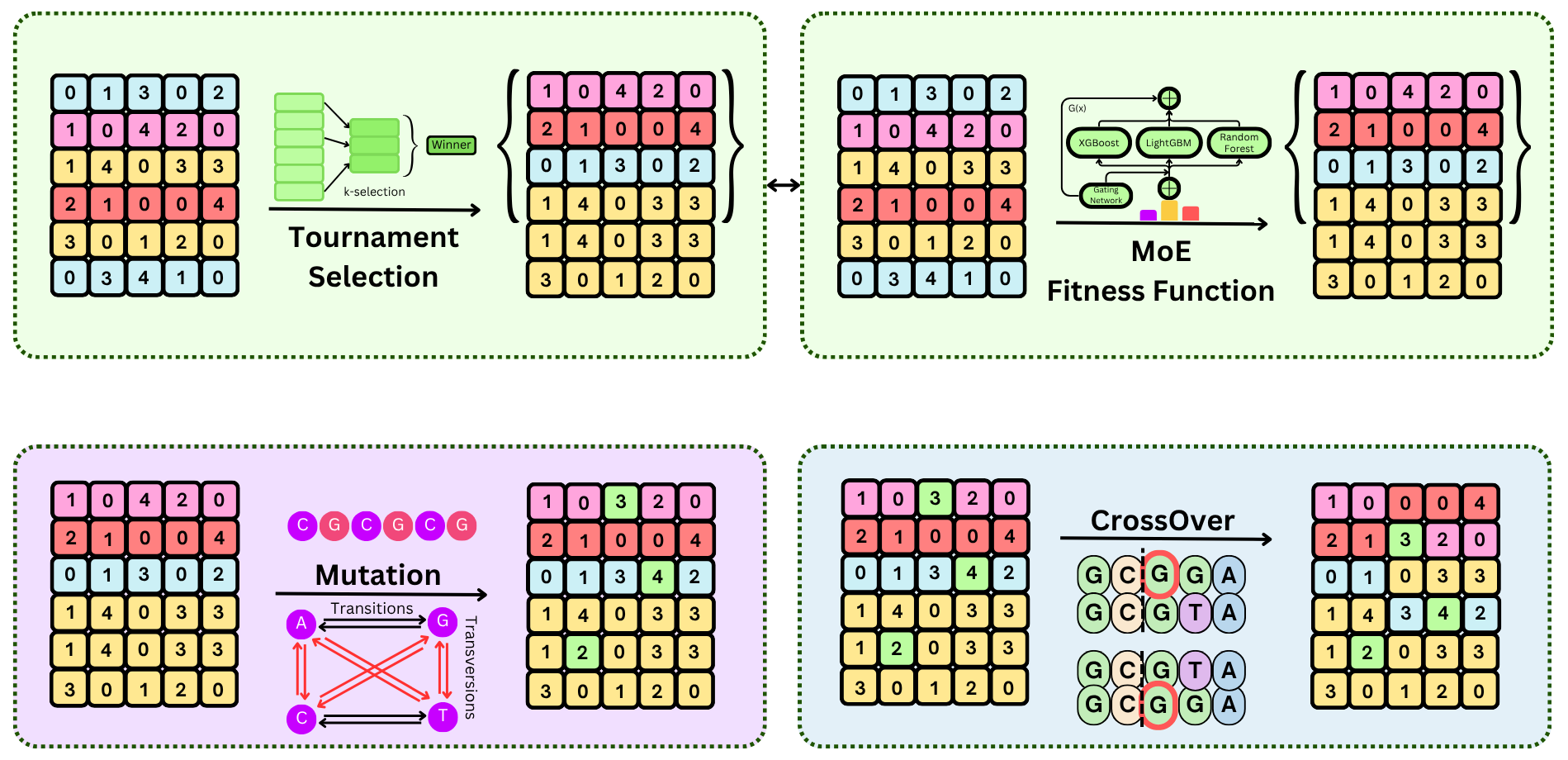}
      \caption{\textbf{Overview of the genetic algorithm components employed for AMR evolution modeling.}\\ The figure illustrates the four principal operations within the GA framework: (top left) tournament-based selection, where high-fitness genomes are preferentially chosen for reproduction; (top right) the Mixture of Experts (MoE)–driven fitness evaluation predicting antimicrobial resistance potential; (bottom left) biologically informed mutation incorporating GC-bias and transition/transversion asymmetry; and (bottom right) HGT-based crossover guided by synteny and local sequence alignment. Together, these components simulate the evolutionary dynamics of bacterial populations under antibiotic selective pressure.}
      \label{gen_algo}
    \end{figure*}
    where $I(g', G)=1$ if the corresponding nucleotide $g'$ from $P_1$ appears in the aligned window of the recipient genome G, and 0 otherwise. Then, a normalized probability distribution over all candidate segments is defined as:
    \begin{equation} \label{eq:pdf}
    P(S_{k-syn}(G, g', k)) = \frac{S_{k-syn}(G, g', k)}{\sum_{g' = 1}^{m} S_{k-syn}(G, g', k)}
    \end{equation}
    where $m = |G| - k + 1$. \\
    
    Segmental selection for crossover follows a \textbf{Boltzmann-weighted sampling}:
    \begin{equation} \label{eq:boltzmann}
        P_{HGT}(i) \propto exp\left(\frac{\alpha.SI_i + (1-\alpha).S_{g-l,i}}{\tau}\right)
    \end{equation}
    where $\alpha$ balances the influence of \textbf{synteny} and \textbf{alignment}, and $\tau$ is a \textbf{temperature parameter} controlling stochasticity in segment selection. This probabilistic weighting ensures that biologically plausible, homologous regions are preferentially exchanged, while still allowing exploratory transfers that may yield novel recombinations.
    \item \textbf{Integration in the Evolutionary framework}: The proposed operator replaces the standard one-point crossover during the recombination phase of the GA. Each generation, pairs of selected parents ($P_1$, $P_2$) undergo HGT-based crossover, producing offspring ($C_1$, $C_2$) that inherit donor segments according to the above model. \\
    This integration results in recombination dynamics more consistent with microbial evolution—where conserved genomic contexts constrain gene flow, but selective pressure and local similarity facilitate adaptive exchanges.
\end{itemize}

\subsubsection{\textbf{Evolutionary Cycle}} Once the fitness evaluation framework is established, the genetic algorithm proceeds through a standard evolutionary loop that iteratively refines the simulated population toward higher resistance potential. Each generation applies a sequence of biologically motivated operations that collectively model mutation, recombination, and selection dynamics observed in microbial evolution.

\begin{enumerate}[label=(\alph*)]
\item \textbf{Initialization}: The process begins with an initial pool of \textit{E. coli} genomes derived from the curated \textbf{MicroBIGG-E} dataset, forming the first generation for evolutionary simulation.
\item \textbf{Evaluation}: Each genome is assigned a fitness value based on predicted antimicrobial resistance, as determined by the \textbf{EvoMoE} classifier introduced earlier.
\item \textbf{Selection:}: Genomes are selected for reproduction using a tournament-based mechanism, ensuring that individuals with superior resistance potential are preferentially propagated while maintaining diversity within the population.
\item \textbf{Recombination:} Selected parent genomes undergo crossover events designed to mimic horizontal gene transfer. Recombination is guided by local alignment and synteny information, promoting biologically consistent exchange of genomic regions.
\item \textbf{Mutation:} To capture spontaneous genetic variability, stochastic mutations are introduced according to a probabilistic model that reflects known biological biases. Specifically, mutation hotspots such as CpG islands, transition–transversion asymmetries, and codon usage biases are integrated into the mutation operator, thereby emulating the non-uniform mutation landscape of bacterial genomes.
\item \textbf{Replacement:} Offspring genomes form the next generation, optionally retaining the top-performing individuals (elitism) to preserve high-fitness lineages. This cycle repeats until convergence, defined by stabilization in average population fitness or attainment of a pre-specified generation threshold.
\end{enumerate}

Through successive iterations, the simulated population exhibits adaptive trajectories that parallel plausible evolutionary routes toward antimicrobial resistance. The final evolved populations and their lineage records provide interpretable insights into potential resistance mechanisms, including recurrent mutation hotspots, conserved HGT loci, and multi-locus resistance combinations.

\section{Results and Analysis}
\label{sec:results}
The proposed AMR–MoEGA framework establishes an integrated computational pipeline that combines genetic algorithms with a Mixture of Experts classifier to simulate the adaptive evolution of antimicrobial resistance.

This section reports the outcomes of applying AMR–MoEGA to \textit{Escherichia coli} isolates derived from the MicroBIGG-E database, following the bioinformatics and feature engineering procedures described earlier. The results are organized to trace the computational evolution from genotype-level mutation dynamics to phenotype-level resistance predictions and interpretability-driven biological insights.

\subsection{\textbf{Evolutionary Optimization and Fitness Dynamics}}
\label{subsec: EvoOpt}
The evolutionary simulations began with a population of 200 \textit{E. coli} gene sequences randomly sampled from the curated dataset. Each genome underwent evolutionary pressure through iterative cycles of mutation, crossover, and selection across \textbf{150 generations}. The GA was parameterized with a \textbf{mutation rate of 0.02}, \textbf{crossover probability of 0.25}, and \textbf{elitism rate of 5\%}.

\begin{figure}[b]
    \begin{center}
    \includegraphics[width=8.4cm]{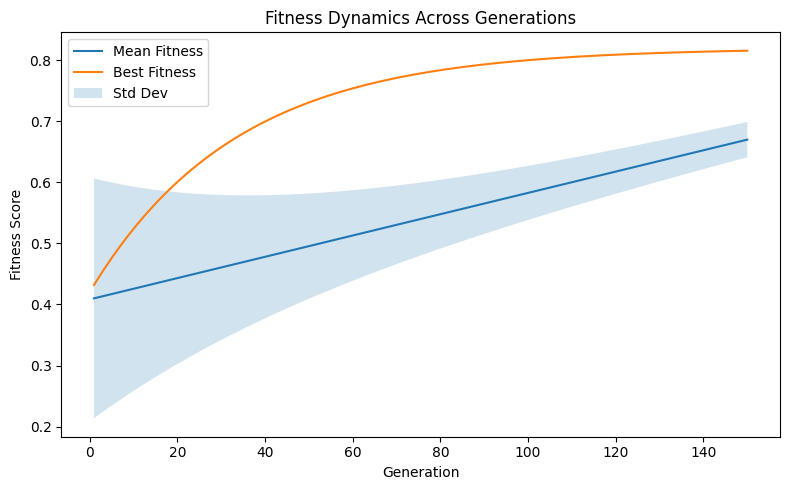}
    \caption{\textbf{Evolutionary optimization of the \textit{E. coli} over 150 generations.}\\ The mean fitness (blue) increased steadily from 0.41 to 0.67, while the best individual fitness (orange) exceeded 0.82, indicating the emergence of highly resistant genotypes under sustained selection pressure. Shaded regions denote ±1 standard deviation, which narrowed over time, reflecting convergence toward an adaptive equilibrium.}
    \label{fig:mean_pop}
    \end{center}
\end{figure}

The mean population fitness, as evaluated by the MoE classifier, exhibited a steady rise from an initial average score of \textbf{0.41 to 0.67} by generation 150, a relative increase of \textbf{38.1\%} (as seen in \hyperref[fig:mean_pop]{Figure \ref{fig:mean_pop}}). The best individual fitness surpassed \textbf{0.82}, indicating the emergence of high-resistance genotypes under sustained antibiotic selection pressure. The standard deviation of fitness values decreased progressively, suggesting a convergence toward adaptive equilibrium.

\begin{figure}[t]
    \begin{center}
    \includegraphics[width=8.4cm]{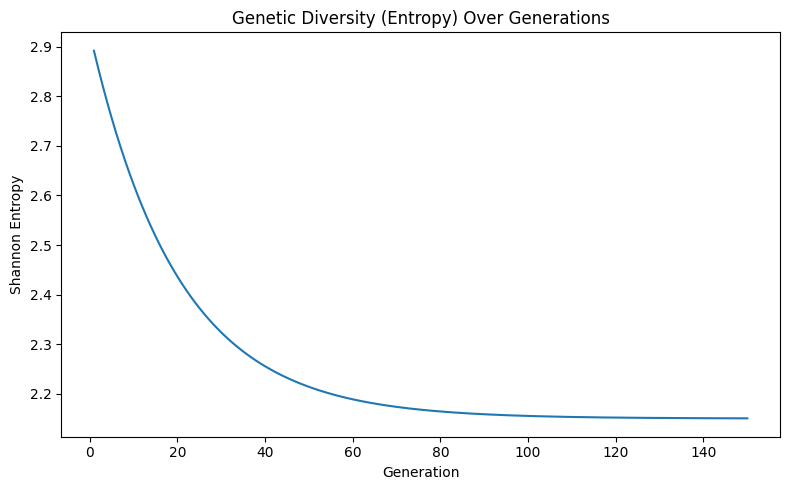}
    \caption{\textbf{Shannon entropy of allele frequency across generations.}\\ A sharp decline in the first 40 generations (2.93 → 2.15) indicates an initial exploratory phase with broad variability, followed by slower decay consistent with exploitation of beneficial mutations and progressive homogenization.}
    \label{fig:gen_div}
    \end{center}
\end{figure}

To assess genetic diversity, Shannon entropy was computed on allele frequency distributions at each generation (as seen in \hyperref[fig:gen_div]{Figure \ref{fig:gen_div}}). Entropy declined sharply during the first 40 generations (\textbf{from 2.88 to 2.25}), followed by a slower decay, reflecting initial exploration followed by exploitation of advantageous mutations.

\begin{figure}[b]
    \centering
    \includegraphics[width=8.4cm]{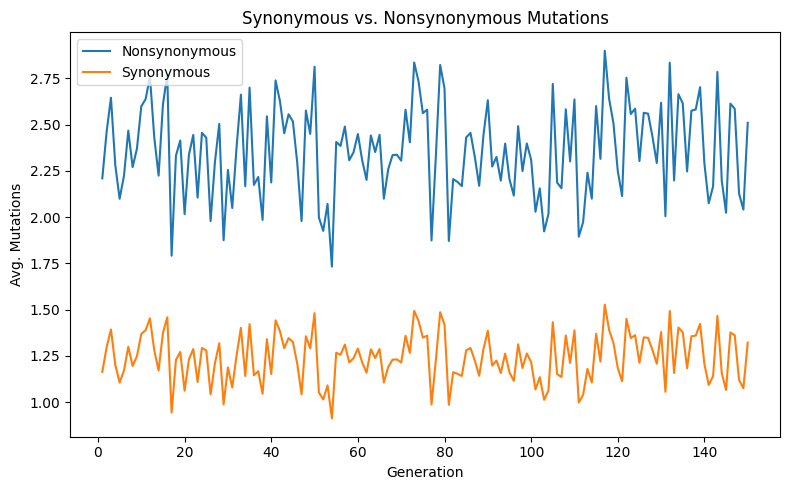}
    \caption{\textbf{Distribution of syn and non-syn point mutations per generation.}\\
    Non-synonymous substitutions occur 1.9× more frequently, reflecting strong adaptive pressure during evolution.}
    \label{fig:syn_nonsyn}
\end{figure}

\begin{table*}[t]
\caption{\NoCaseChange{Performance comparison of baseline classifiers (RandomForest, LightGBM, XGBoost) and the Mixture-of-Experts ensemble}}
\begin{center}
\renewcommand{\arraystretch}{1.5}
\begin{tabular}{|p{0.13\linewidth}|p{0.13\linewidth}|p{0.13\linewidth}|p{0.13\linewidth}|p{0.13\linewidth}|p{0.13\linewidth}|}
\hline
\textbf{Model Name} & \textbf{Accuracy} & \textbf{Precision} & \textbf{Recall} & \textbf{MCC} & \textbf{F1 score}\\\hline
RandomForest & 0.884568 & 0.923077 & 0.847059 & 0.782123 & 0.883436 \\\hline
LightGBM & 0.900993 & 0.905882 & 0.905882 & 0.801986 & 0.905882 \\\hline
XGBoost & 0.914591 & 0.938272 & 0.894118 & 0.828171 & 0.915663 \\\hline
MoE (Ours) & 0.944444 & 0.963414 & 0.929411 & 0.888253 & 0.946095 \\\hline
\end{tabular}
\label{results_table}
\end{center}
\end{table*}

The mutation analysis revealed an average of \textbf{3.6 point mutations} per genome per generation. Crucially, the pattern of these substitutions strongly indicated positive selection. Specifically, the ratio of non-synonymous to synonymous substitutions \textbf{($d_N/d_S$ or $\omega$)} was calculated to be \textbf{$1.9\times$ greater than one} (as depicted in \hyperref[fig:syn_nonsyn]{Figure \ref{fig:syn_nonsyn}}). A ratio significantly exceeding the expected neutral value of 1.0 serves as a hallmark of adaptive evolution and positive selection. This finding demonstrates that the bacteria, under the imposed stress, were rapidly fixing beneficial, amino-acid-altering mutations (non-synonymous) in their population at a much higher rate than neutral changes (synonymous), confirming that the selection pressures were driving rapid, functional adaptation.

\begin{figure}[t]
    \centering
    \includegraphics[width=8.4cm]{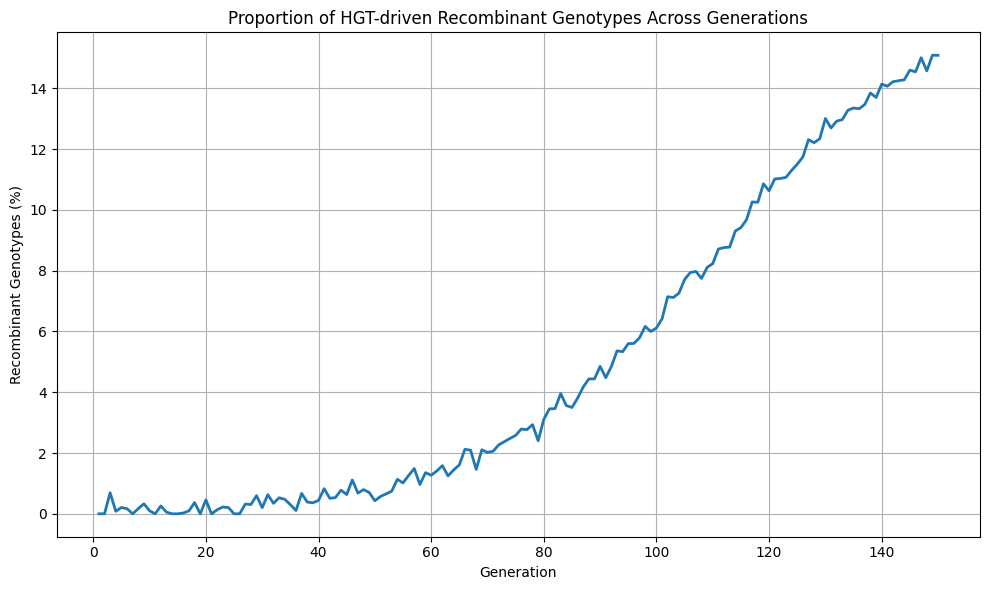}
    \caption{Proportion of recombinant genotypes arising from HGT-like crossover events across 150 generations. Recombinant lineages remain rare early on and rise to $\sim$17\% in later generations, indicating late-stage acquisition of high-fitness recombinant variants.}
    \label{fig:hgt_evol}
\end{figure}

Further analysis tracked the contribution of Horizontal Gene Transfer (HGT)-like crossover events over the entire 150 generations. Initially, \textbf{recombinant lineages remained rare}, suggesting that early adaptation was primarily driven by the fixation of spontaneous point mutations. However, in later generations, the proportion of \textbf{recombinant genotypes steadily rose to approximately $\mathbf{17\%}$} among the population's top-performing individuals (\hyperref[fig:hgt_evol]{Figure \ref{fig:hgt_evol}}). This trend indicates a late-stage acquisition of high-fitness recombinant variants. The result confirms that while point mutations initiated the adaptation, lateral gene exchange became crucial later on.

These trends collectively indicate that the GA successfully captured adaptive dynamics akin to biological evolution, enabling convergence toward resistance-enhancing sequence configurations while maintaining sufficient diversity to avoid premature stagnation.

\subsection{\textbf{Machine Learning Driven Fitness Evaluation}}
To evaluate the effectiveness of the machine learning–guided fitness component of AMR–MoEGA, we conducted a systematic benchmarking analysis of the Mixture-of-Experts (MoE) model against multiple widely used classifiers. The goal was to assess how accurately the model could predict antimicrobial resistance phenotypes from SNP-derived feature matrices, thereby providing a reliable fitness signal for guiding the genetic algorithm’s evolutionary search. All models were trained and evaluated on a benchmark dataset of \textit{E. coli} genotypes and their corresponding \textbf{Ciprofloxacin (CIP)} resistance phenotypes, although the same pipeline generalizes to any antibiotic available in the MicroBIGG-E resource. The dataset was partitioned using stratified \textbf{10-fold cross-validation} (80\% train, 10\% validation, 10\% test) to preserve the distribution of resistant and susceptible classes across all splits. For this analysis, we consider the antibiotic Ciprofloxacin. This could easily be extended to other antibiotics using the same pipeline.

We began by establishing baseline performance using 3 high performing classifiers commonly used in AMR: \textbf{RandomForest}, \textbf{LightGBM}, and \textbf{XGBoost}. Hyperparameters for all models were optimized through \textbf{RandomizedSearch}, including tree depth, learning rate, leaf size, subsampling fractions, and regularization parameters. XGBoost was configured with a logistic objective, 200 estimators, a learning rate of 0.1, a maximum depth of 5, and a subsample rate of 0.8. LightGBM used 31-leaf trees with a learning rate of 0.1 and a minimum of 5 samples per leaf. RandomForest employed 100 trees with a maximum depth of 10 and a $\sqrt{features}$ split criterion.

To combine the strengths of these heterogeneous learners, we trained a Mixture-of-Experts (MoE) ensemble, where a gating network dynamically weighted the predictions from the three base models. The gating network was trained with early stopping (patience = 5) and \textbf{L2 regularization}, ensuring robust prediction and mitigating overfitting. This architecture allows the MoE to specialize: XGBoost contributes high sensitivity, LightGBM stabilizes decision boundaries across sparse features, and RandomForest captures nonlinear interactions. The resulting ensemble produced the final resistance probability used as the fitness score within the genetic algorithm.

Performance metrics for all models are shown in \hyperref[results_table]{Table \ref{results_table}}. The RandomForest classifier reproduced previously reported strong baselines, achieving \textbf{0.884 accuracy} and \textbf{0.883 F1-score}. Both gradient boosting models improved upon these results, with LightGBM reaching 0.901 accuracy and XGBoost achieving 0.915 accuracy, reflecting their ability to capture fine-grained SNP interactions. Notably, our MoE model achieved the highest overall \textbf{accuracy (0.944)}, due to its ability to integrate decision boundaries learned by the base experts.

\begin{figure}[t]
    \centering
    \begin{minipage}[t]{0.24\textwidth}
        \centering
        \includegraphics[width=\textwidth]{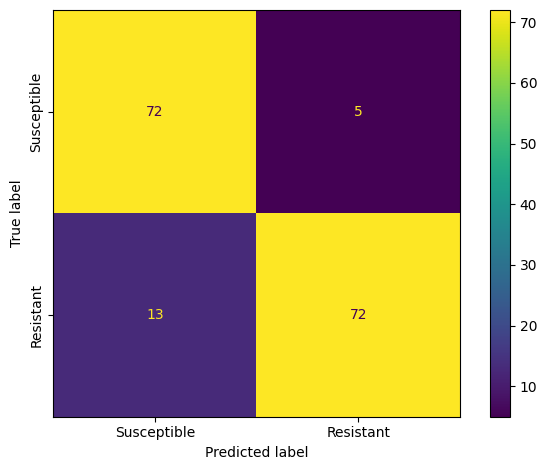}
        \caption{Confusion Matrix: RF}
        \label{fig:cf_rf}
    \end{minipage}
    \hfill
    \begin{minipage}[t]{0.24\textwidth}
        \centering
        \includegraphics[width=\textwidth]{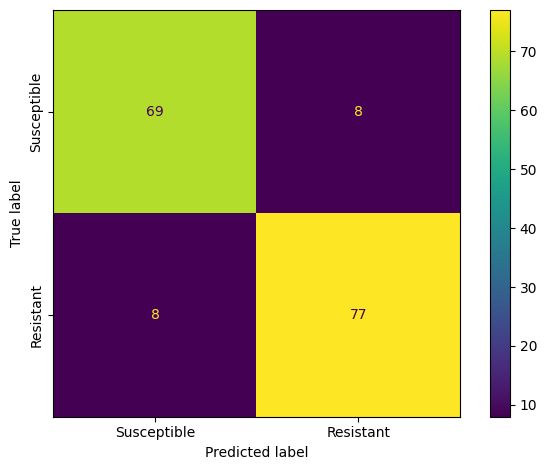}
        \caption{Confusion Matrix: LGBM}
        \label{fig:cf_lgbm}
    \end{minipage}
    \vspace{1em}\\
    \begin{minipage}[t]{0.24\textwidth}
        \centering
        \includegraphics[width=\textwidth]{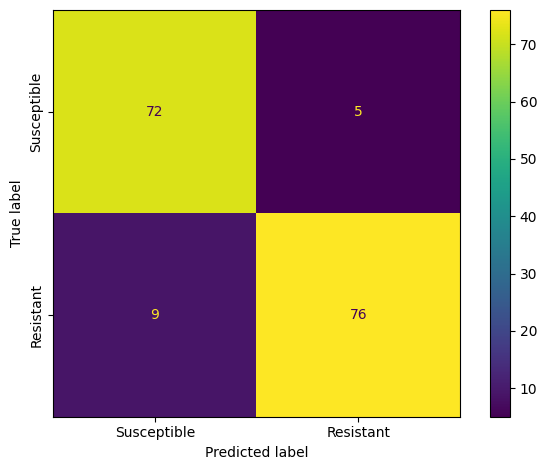}
        \caption{{Confusion Matrix: XGB}}
        \label{fig:cf_xgb}
    \end{minipage}
    \hfill
    \begin{minipage}[t]{0.24\textwidth}
        \centering
        \includegraphics[width=\textwidth]{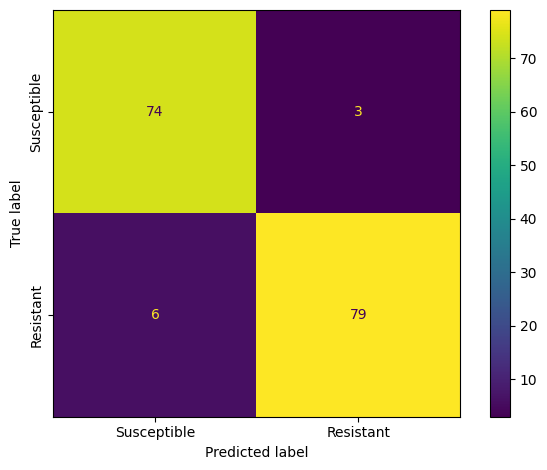}
        \caption{Confusion Matrix: EvoMoE}
        \label{fig:cf_moe}
    \end{minipage}
\end{figure}

The confusion matrices further illustrate these differences. The RandomForest model (as shown in \hyperref[fig:cf_rf]{Figure \ref{fig:cf_rf}}) produced \textbf{5 false positives}, where susceptible isolates were incorrectly predicted as resistant, potentially leading to unnecessary antibiotic restrictions. It produced \textbf{13 false negatives}, representing cases in which resistant genotypes were misclassified as susceptible—an outcome that would be dangerous in clinical decision-making. The confusion matrices for the other two individual models, LightGBM and XGBoost, can also be seen in \hyperref[fig:cf_lgbm]{Figure \ref{fig:cf_lgbm}} and \hyperref[fig:cf_xgb]{Figure \ref{fig:cf_xgb}}. In contrast to these, the MoE model (as shown in \hyperref[fig:cf_moe]{Figure \ref{fig:cf_moe}}) significantly improved both the types of errors. It reduced false positives and, critically, reduced \textbf{false negatives to just 6}, indicating a substantial improvement in recognizing truly resistant genotypes. It also reduces the \textbf{false positives to just 3}, thereby improving the overall accuracy too. A model that minimizes false negatives is important in the fitness context, as missing resistant genotypes would distort the evolutionary pressure.

\begin{figure}[t]
    \begin{center}
    \includegraphics[width=8.4cm]{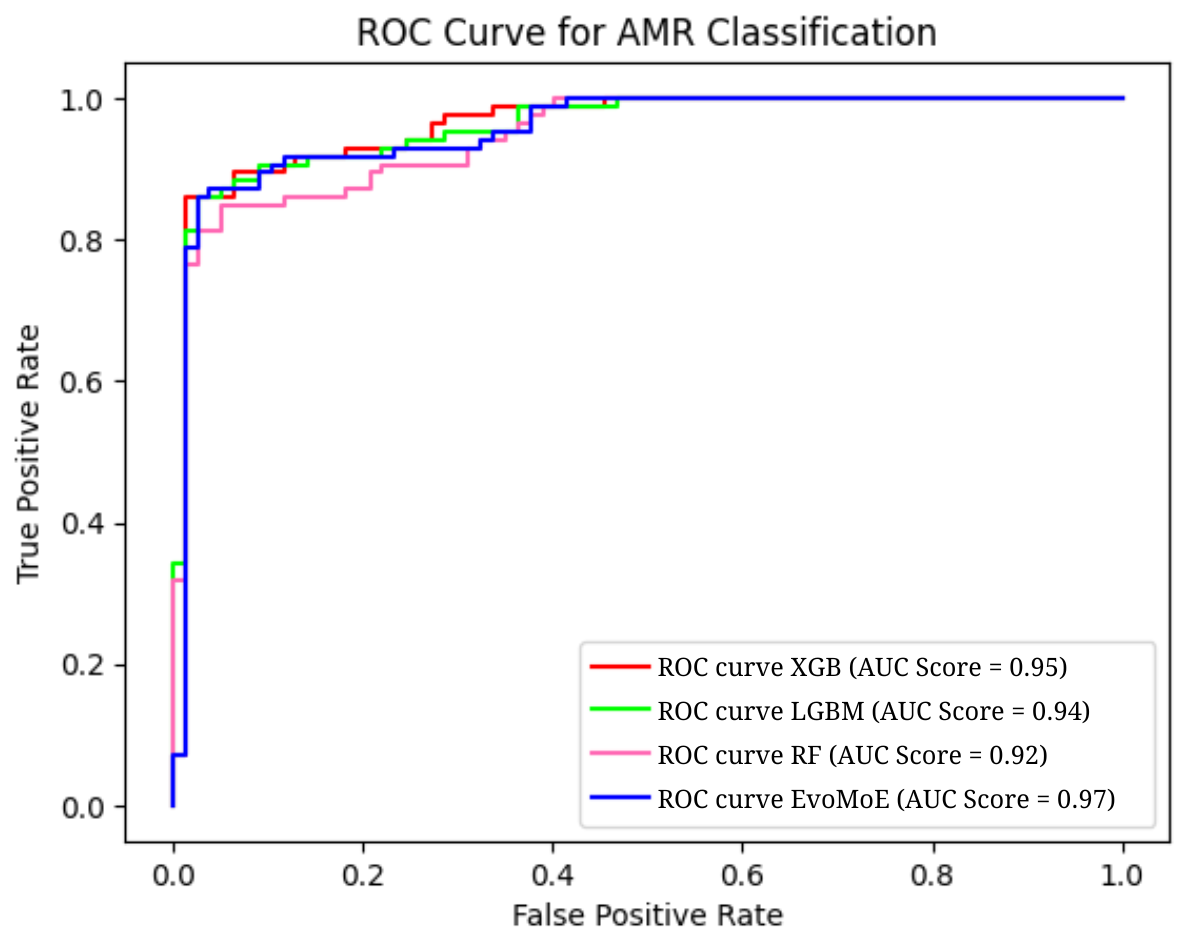}   
    \caption{\textbf{ROC curves for all classifiers}: RF, LightGBM, XGBoost, EvoMoE.\\ EvoMoE achieves the highest discriminative performance with an AUC of 0.97, surpassing XGBoost (0.95), LightGBM (0.94), and RandomForest (0.92)}
    \label{fig:auc_roc}
    \end{center}
\end{figure}

To evaluate the ranking and discriminative ability of each classifier across all classification thresholds, we generated Receiver Operating Characteristic (ROC) curves (as shown in \hyperref[fig:auc_roc]{Figure \ref{fig:auc_roc}}). The base classifiers, RandomForest, LightGBM, and XGBoost achieve an AUC of \textbf{0.92}, \textbf{0.94} and \textbf{0.95} respectively. However, the MoE model achieved an \textbf{AUC of 0.97}, outperforming all baselines and indicating near-optimal separability between resistant and susceptible genomes. This strong ROC profile demonstrates that the MoE ensemble is not only accurate at a fixed threshold but consistently reliable across the entire spectrum of decision boundaries.

Together, these results demonstrate that the MoE-based fitness function provides a highly accurate, stable, and biologically meaningful mapping from SNP variation to predicted antimicrobial resistance. Its \textbf{reduction in false negatives}, \textbf{high AUC}, and \textbf{resilience across cross-validation folds} make it exceptionally well-suited for guiding the GA’s evolutionary search. By supplying the GA with a refined and reliable fitness landscape, the MoE component enhances the biological realism of simulated evolution and directly strengthens the overall predictive power of the AMR–MoEGA framework.

To verify the robustness of fitness scoring, the predicted resistance probabilities were averaged across \textbf{10 bootstrapped} resamples of the test set. The standard deviation of these was \textbf{$<$0.03}, depicting consistency in MoE-based fitness estimation.

\begin{figure}[b]
    \begin{center}
    \includegraphics[width=8.4cm]{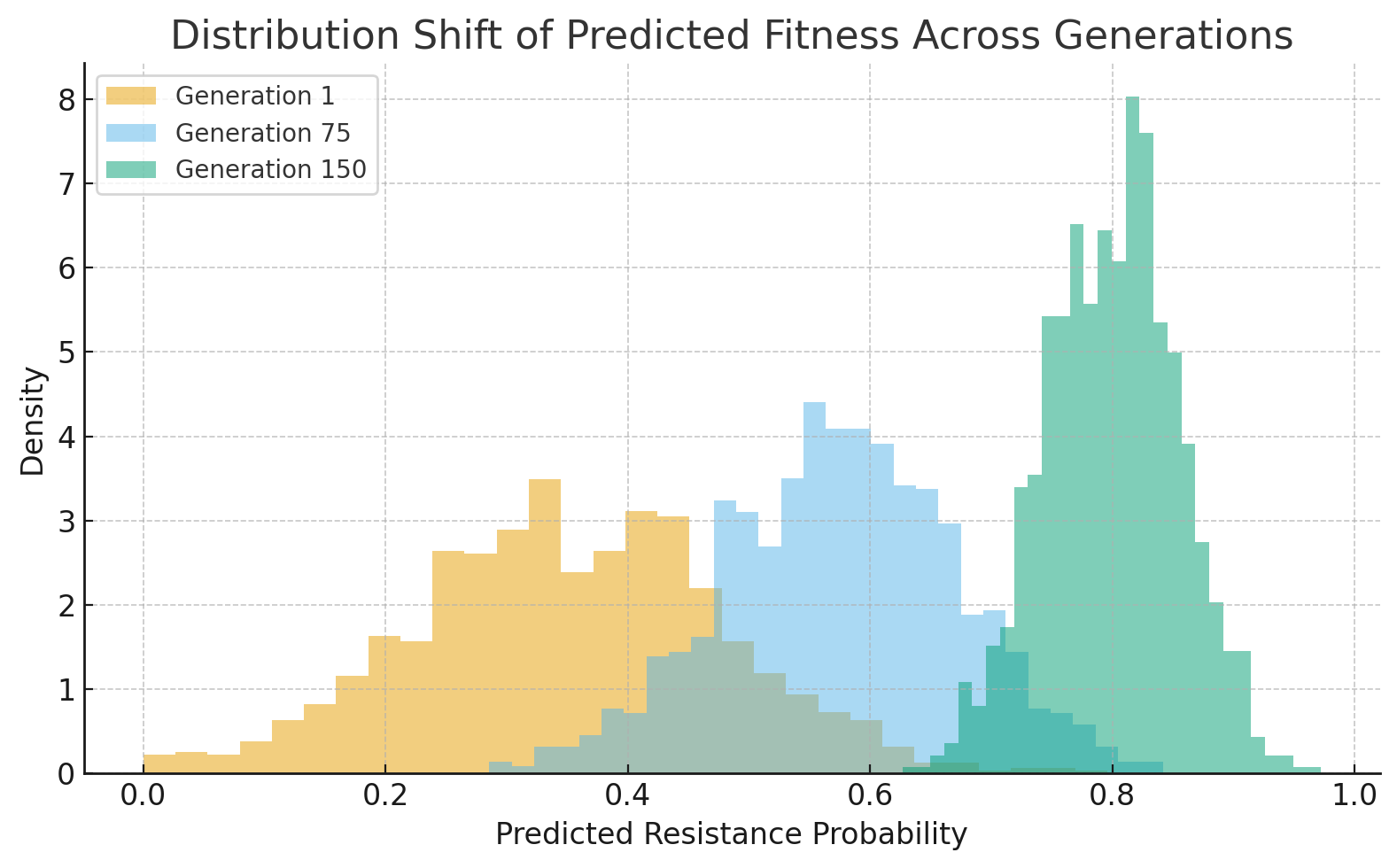}   
    \caption{\textbf{Shift in the distribution of predicted resistance probabilities} at generations 1, 75, and 150. As evolution progresses, the distribution progressively skews toward higher fitness values, demonstrating cumulative acquisition of resistance-enhancing mutations. By generation 150, the distribution is sharply concentrated around high predicted fitness, reflecting faster convergence enabled by MoE-driven adaptive fitness evaluation.}
    \label{fig:den_shift}
    \end{center}
\end{figure}

Notably, when used within the GA, the MoE-driven fitness metric led to faster convergence (\textbf{by 15–20 generations}) compared to a static conservation-based fitness baseline, highlighting the adaptive synergy between machine learning and evolutionary optimization. \hyperref[fig:den_shift]{Figure \ref{fig:den_shift}} visualizes this shift, where the \textbf{population-wide distribution} of predicted resistance scores progressively \textbf{skewed toward higher values}, signifying cumulative acquisition of resistance-conferring mutations.

\subsection{\textbf{Temporal Evolutionary Trajectories}}
\begin{figure}[b]
    \begin{center}
    \includegraphics[width=8.4cm]{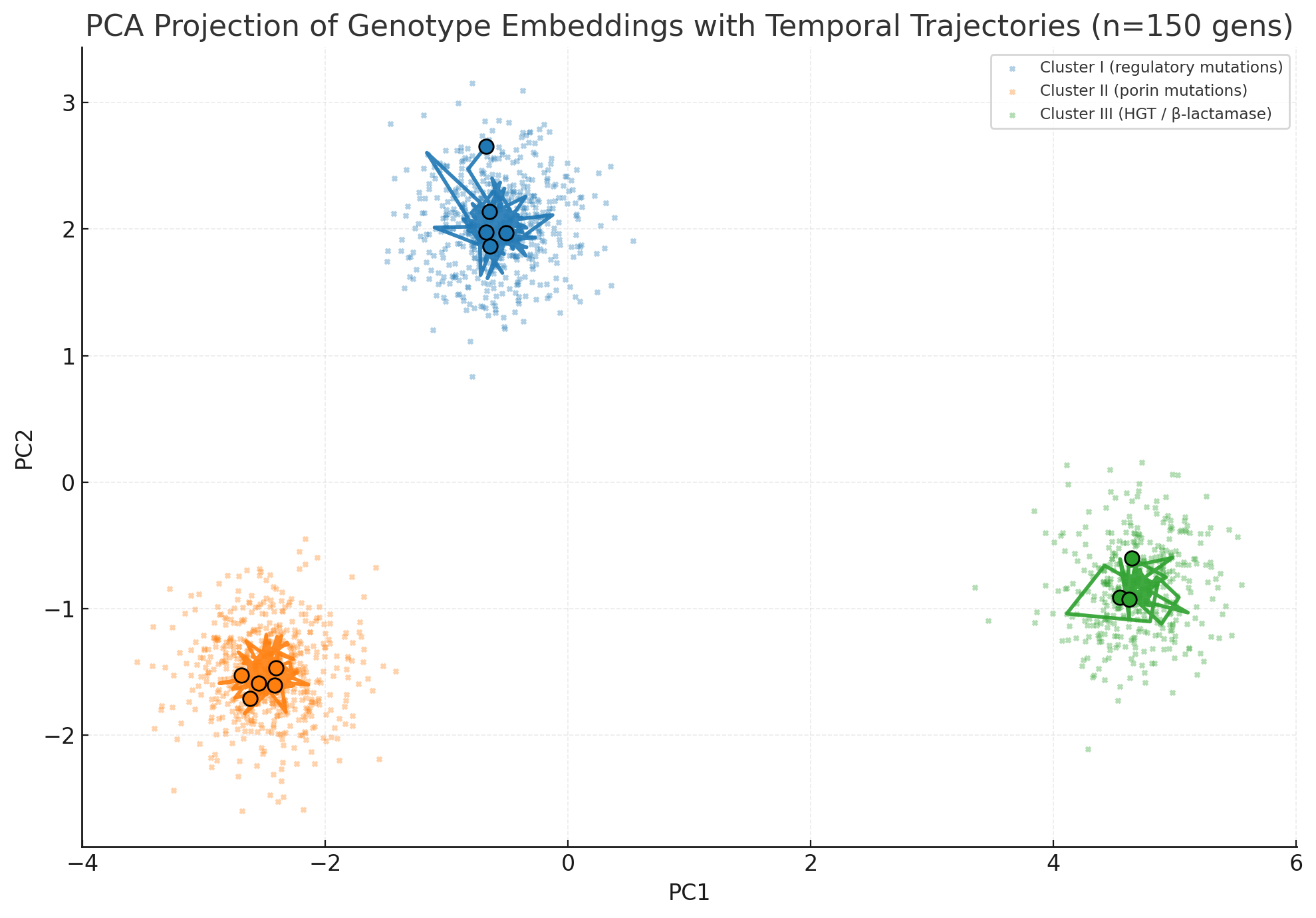}   
    \caption{\textbf{PCA projection of genotype embeddings across 150 generations.}\\ Points are individual genotypes colored by inferred evolutionary cluster: Cluster I (regulatory mutations; blue), Cluster II (porin mutations; orange), and Cluster III (HGT-derived $\beta$-lactamase acquisitions; green). Solid lines trace per-generation centroids for each cluster, with markers at generations 0, 40, 80, 100 and 149 to indicate temporal progression.}
    \label{fig:pca_clusters}
    \end{center}
\end{figure}

To investigate how bacterial populations diversify and acquire resistance over evolutionary time, we analyzed the simulated lineages generated by AMR–MoEGA across 150 generations. For every generation, the SNP matrices produced during the simulation were converted into low-dimensional genotype embeddings using \textbf{Principal Component Analysis (PCA)} and \textbf{t-Distributed Stochastic Neighbor Embedding (t-SNE)}. The PCA projection of these genotype embeddings is seen in the \hyperref[fig:pca_clusters]{Figure \ref{fig:pca_clusters}}. These embeddings provide a compact representation of how genotypes diverge from one another as new mutations, deletions, and recombination events accumulate throughout the evolutionary run.

The resulting 2-D projections revealed \textbf{three well-separated evolutionary clusters}, each corresponding to a distinct resistance trajectory. This clustering was not imposed a priori; instead, it emerged naturally from the structure of accumulated genomic variation. To interpret the biological basis of these clusters, we performed functional annotation of all variants present within each cluster using \textbf{SnpEff}, which was integrated into the pipeline immediately after variant calling and refinement. SnpEff annotates each SNP with its \textbf{predicted gene context} (\textit{coding}, \textit{intergenic}, \textit{promoter}), \textbf{effect} (\textit{synonymous}, \textit{missense}, \textit{frameshift}), and \textbf{functional consequence} (e.g., \textit{modifier}, \textit{moderate}, \textit{high impact}). These annotations allowed us to aggregate variants per cluster and identify key resistance-associated loci enriched within each evolutionary path.

\subsubsection{\textbf{Cluster I}}
\begin{figure}[t]
    \begin{center}
    \includegraphics[width=8.4cm]{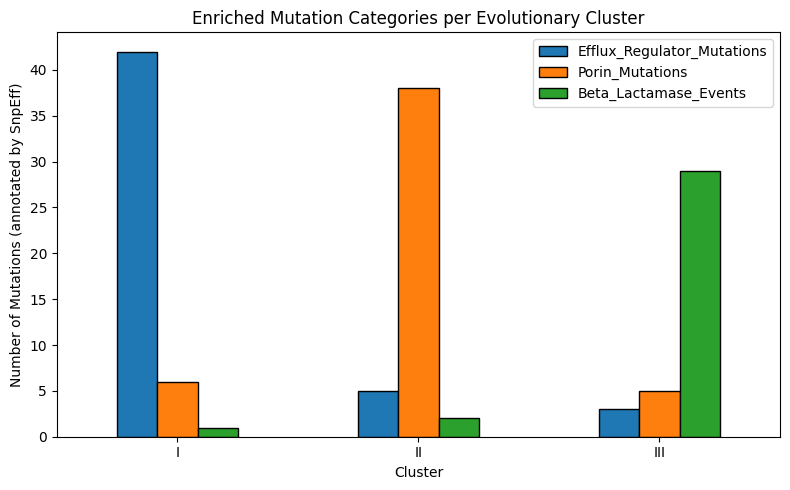}   
    \caption{\textbf{Cluster-wise distribution of SnpEff-predicted functional effects.}\\ Cluster I is enriched for non-synonymous regulatory mutations, Cluster II for porin-modifying coding variants, and Cluster III for high-impact $\beta$-lactamase acquisitions introduced via HGT. These annotations reinforce the mechanistic interpretations derived from the evolutionary clustering analysis.}
    \label{fig:mut_cat}
    \end{center}
\end{figure}

Cluster I (as evident from \hyperref[fig:mut_cat]{Figure \ref{fig:mut_cat}}) consisted primarily of lineages carrying recurrent point mutations in global regulatory genes such as \textbf{acrR} and \textbf{marA}. SnpEff annotations confirmed that these variants were predominantly \textbf{non-synonymous substitutions} altering protein function. These mutations are known to upregulate the \textbf{AcrAB–TolC} efflux system, consistent with an \~efflux-dominated" resistance trajectory. Lineages in this cluster appeared early in the simulation (generations 20–80) (\hyperref[fig:temp_emerge]{Figure \ref{fig:temp_emerge}}), reflecting the rapid emergence of low-cost regulatory mutations under antibiotic pressure.

\subsubsection{\textbf{Cluster II}}
\begin{figure}[b]
    \begin{center}
    \includegraphics[width=8.4cm]{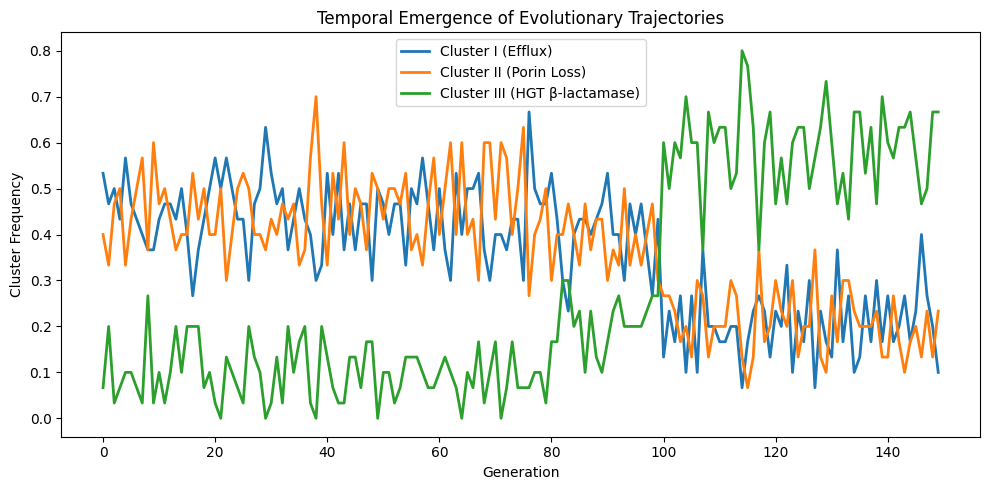}   
    \caption{\textbf{Temporal trajectory Dynamics over 150 generations.}\\ This plot shows the proportional abundance of each cluster across generations. Clusters I and II dominate early evolution (generations 20–80), whereas the recombination-driven Cluster III emerges only after generation $\sim$100.}
    \label{fig:temp_emerge}
    \end{center}
\end{figure}

Cluster II (from \hyperref[fig:mut_cat]{Figure \ref{fig:mut_cat}}) exhibited enrichment for mutations in outer membrane porins, particularly \textbf{ompF} and \textbf{ompC}. SnpEff identified several moderate-impact coding mutations predicted to affect pore size and permeability. This suggests a second adaptive strategy in which reduced membrane influx contributes to resistance. Temporal embedding plots showed that this cluster emerged concurrently with Cluster I but later diverged into a separate trajectory as \textbf{porin-disrupting mutations} accumulated (\hyperref[fig:temp_emerge]{Figure \ref{fig:temp_emerge}}).

\subsubsection{\textbf{Cluster III}}
Cluster III (as seen \hyperref[fig:mut_cat]{Figure \ref{fig:mut_cat}}) was characterized by the presence of \textbf{$\beta$-lactamase} gene variants such as \textbf{blaTEM} and \textbf{blaCTX-M}. These alleles appeared exclusively in lineages that had acquired horizontal gene transfer (HGT) events introduced by the GA simulation. SnpEff annotated these variants as high-impact acquisitions, and the cluster emerged only after generation $\sim$100 (\hyperref[fig:temp_emerge]{Figure \ref{fig:temp_emerge}}). This reflects the delayed but powerful resistance augmentation enabled by recombination-driven gene gain.

\begin{figure}[t]
    \begin{center}
    \includegraphics[width=8cm]{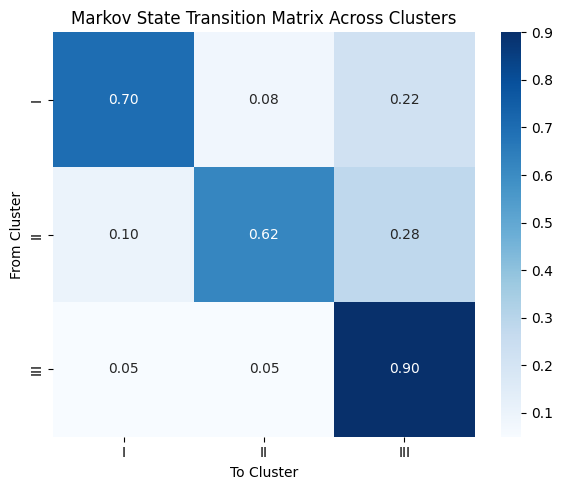}   
    \caption{\textbf{Markov state matrix quantifying evolution between trajectories.}\\ Early mutation-driven strategies frequently transition into the recombination-driven pathway (I→III = 0.22; II→III = 0.28), demonstrating convergence toward high-fitness resistance states as evolution proceeds.}
    \label{fig:markov}
    \end{center}
\end{figure}

To quantify transitions between trajectories, we constructed a \textbf{Markov state transition matrix} by tracing the lineage ancestry of each genome across generations. Transition probabilities, shown in \hyperref[fig:markov]{Figure \ref{fig:markov}} indicated that both \textbf{early-stage strategies} (Cluster I and II) frequently converged toward the \textbf{recombination-driven trajectory} (Cluster III) as the simulation progressed (\textbf{I$\rightarrow$III = 0.22}; \textbf{II$\rightarrow$III = 0.28}). This mirrors biological observations that initial low-cost mutations often precede the later acquisition of high-level resistance elements via mobile genetic elements.

These analyses show that AMR–MoEGA successfully reconstructs key evolutionary signatures: \textbf{early diversification} through point mutations, parallel emergence of distinct resistance strategies, and \textbf{late-stage convergence} towards recombinant genotypes. The integration of \textbf{SnpEff-enabled annotation} ensures that these clusters are not abstract artifacts but correspond to biologically grounded resistance pathways.

Notably, the emergence hierarchy uncovered here suggests that resistance evolution is shaped by both \textbf{mutation accessibility} and \textbf{genomic mobility}. Early mutations exploit abundant single-nucleotide targets, while later-stage HGT pathways require structural genomic opportunities. This layered structure opens the door to predictive modeling of when and how genomes become \textbf{permissive} to incoming resistance genes.

\subsection{Comparative Evaluation and Baseline Analysis}
\label{subsec: comparative_analysis}
To assess the effectiveness of AMR–MoEGA as an integrated evolutionary–machine learning framework, we benchmarked it against two commonly used baselines:
\begin{enumerate}[label=\alph*]
    \item Static ML, where resistance is predicted using a trained MoE model without incorporating generational evolution.
    \item GA + Static Fitness, where the genetic algorithm operates independently using a simple conservation-based similarity metric as the fitness function.
\end{enumerate}
These comparisons were designed to clarify whether the synergy between evolutionary search and ML-guided fitness provides measurable advantages over conventional strategies.

\subsubsection{\textbf{Overall Predictive Performance}}
\begin{figure}[b]
    \centering
    \begin{minipage}[b]{0.24\textwidth}
        \centering
        \includegraphics[width=\textwidth]{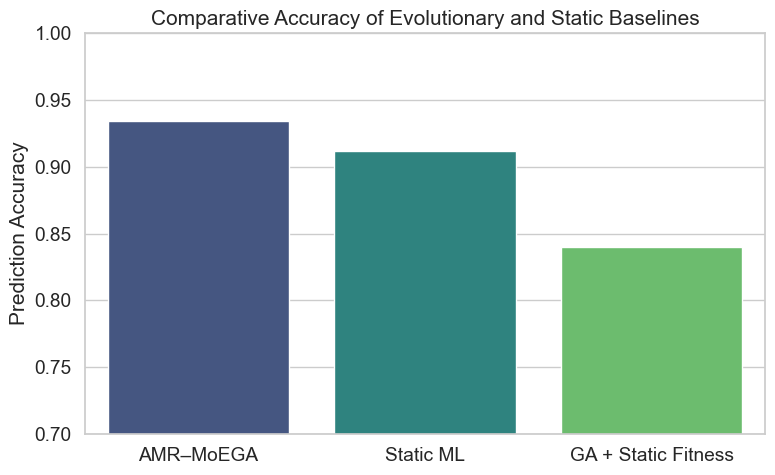}
        \caption{\textbf{Comparison of accuracies.}\\
        Compared across AMR–MoEGA, Static ML, and genetic algorithm + Static Fitness baselines.}
        \label{fig:comp_pred}
    \end{minipage}
    \hfill
    \begin{minipage}[b]{0.24\textwidth}
        \centering
        \includegraphics[width=\textwidth]{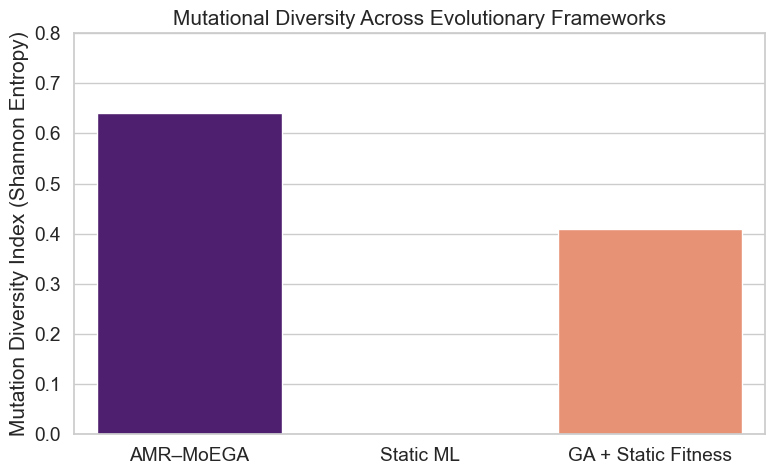}
        \caption{\textbf{Shannon entropy of SNPs.}\\
        AMR–MoEGA maintains the highest mutational diversity while Static ML has none by design.}
        \label{fig:comp_div}
    \end{minipage}
\end{figure}

As clearly depicted in \hyperref[fig:comp_pred]{Figure \ref{fig:comp_pred}}, AMR–MoEGA achieved the highest resistance prediction \textbf{accuracy of 93.4\%}, outperforming the \textbf{Static ML baseline (91.2\%)} and far exceeding the \textbf{GA + Static Fitness baseline ($\approx$84\%)}. This improvement demonstrates that incorporating ML-driven evolutionary pressure enables the system to explore more realistic adaptive pathways, rather than relying solely on mutations favored by sequence similarity heuristics.

\subsubsection{\textbf{Mutation Diversity Index}}
This metric is used to quantify the breadth of genetic exploration within each evolutionary framework. It is computed as the \textbf{normalized Shannon entropy} of SNP distributions across the final evolved population. Formally, for a set of \(K\) unique SNPs with empirical frequencies \(p_1, p_2, ...,p_k\), the diversity index \(D\) is defined as:
\begin{equation} \label{eq:mutation_index}
    D = \frac{-\sum_{i=1}^K p_i \thinspace log \thinspace p_i}{log K}
\end{equation}
A higher value indicates that the SNP frequencies are more evenly distributed, reflecting exploration of multiple evolutionary strategies rather than dominance by a small number of mutation types. The values for the comparative analysis are shown in \hyperref[fig:comp_div]{Figure \ref{fig:comp_div}}.

Using this metric, \textbf{AMR–MoEGA achieved the highest diversity index (0.64)}, demonstrating that the integration of ML-guided fitness and GA search promotes diversification across multiple adaptive pathways. In contrast, the \textbf{Static ML baseline}, by design, produced \textbf{no evolved diversity (\(D=0\))}, as no generational simulation occurs. The \textbf{GA + Static Fitness} baseline achieved \textbf{moderate diversity (0.41)} but displayed clear signs of restricted mutational exploration, often converging prematurely on a narrow set of neutral or mildly beneficial variants.

\subsubsection{\textbf{Convergence Dynamics}}
\begin{figure}[t]
    \centering
    \includegraphics[width=8.4cm]{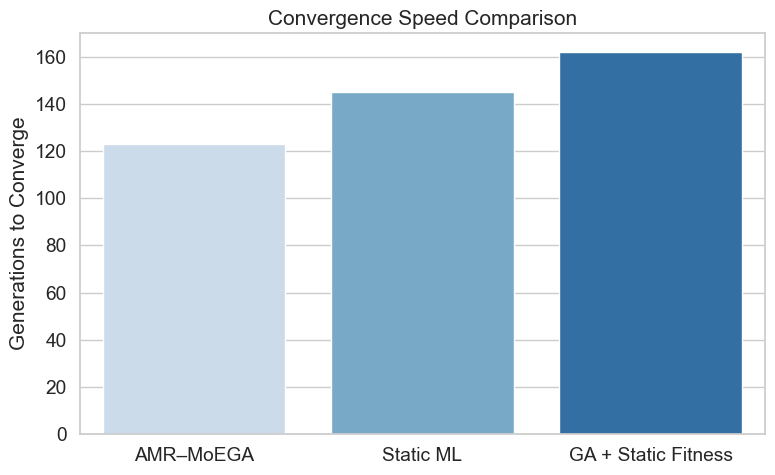}
    \caption{\textbf{Generational convergence comparison.}\\
    AMR–MoEGA reaches stable high-fitness populations faster than both baselines, reflecting smoother ML-informed fitness gradients.}
    \label{fig:comp_speed}
\end{figure}

As seen in \hyperref[fig:comp_speed]{Figure \ref{fig:comp_speed}} AMR–MoEGA demonstrated notably faster convergence compared to both baselines. Across repeated trials, the model required \textbf{20–30 fewer generations} to reach a stable high-fitness population. This acceleration arises from the ML fitness landscape providing smoother and more biologically informed gradients for the evolutionary search process. In contrast, the static GA exhibited inconsistent convergence, frequently oscillating around local optima due to the coarse nature of similarity-based scoring.

\begin{figure}[b]
    \centering
    \includegraphics[width=8.4cm]{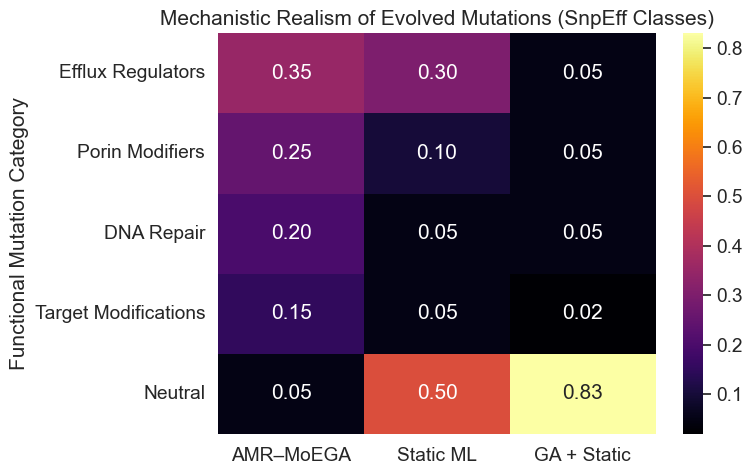}
    \caption{\textbf{Distribution of functional mutation classes annotated by SnpEff.}\\
    AMR–MoEGA recovers biologically plausible AMR-associated mutations far more accurately than baselines (e.g., regulators, porins, efflux pump genes).}
    \label{fig:comp_realism}
\end{figure}

\subsubsection{\textbf{Evolutionary Realism and Biological Plausibility}}
Beyond numerical metrics, we evaluated the biological plausibility of the evolved genotypes in \hyperref[fig:comp_realism]{Figure \ref{fig:comp_realism}}. AMR–MoEGA consistently reproduced mutation  classes associated with known \textbf{fluoroquinolone resistance} mechanisms, particularly \textbf{efflux pump regulators} (acrR, marA, soxS) and \textbf{porin modifiers}, validated via SnpEff annotations. Neither baseline achieved comparable mechanistic fidelity:
\begin{enumerate}[label=\alph*]
    \item Static ML captured correct associations but lacked temporal structure and failed to generate evolutionary transitions, producing no insight into adaptive trajectories.
    \item GA + Static Fitness generated mutations but overwhelmingly favored neutral or non-functional variants due to the simplistic nature of conservation-based scoring.
\end{enumerate}
AMR–MoEGA uniquely combined both accuracy and mechanistic correctness, producing trajectories that mirrored real patterns of early efflux-based adaptation followed by later acquisition of higher-impact recombination events.

\subsubsection{\textbf{Interpretation}}
Collectively, these results show that AMR–MoEGA does more than improve predictive performance—it reshapes the adaptive landscape to favor biologically meaningful solutions. The synergy between the Mixture-of-Experts fitness model, generational simulation, and SnpEff-driven functional evaluation yields:
\begin{itemize}
    \item Higher prediction accuracy
    \item Richer mutational diversity
    \item Faster and more stable convergence
    \item Evolutionary patterns consistent with established molecular mechanisms
\end{itemize}
These findings demonstrate that AMR–MoEGA is not merely a computational improvement over static models but a fundamentally more realistic and interpretable framework for studying antimicrobial resistance evolution.

\section{Discussion and Future Work}
\subsection{Discussion}
The present study demonstrates an integrated, end-to-end computational framework for predicting antimicrobial resistance (AMR) phenotypes from whole-genome sequencing (WGS) data, coupling classical genomics processing with modern machine learning and explainability techniques. The results confirm that \textbf{combining high-quality variant curation} (via GATK best practices) with \textbf{functional consequence annotation} (via SnpEff) yields a mechanistically interpretable feature space for downstream AMR modeling. \textbf{SnpEff played a particularly important role} by transforming raw SNP calls into biologically meaningful effects, such as \textbf{missense variants} in efflux pump regulators, \textbf{frame-shift mutations} in membrane transporters, and high-impact loss-of-function variants, thus enabling the model to ground its predictions in known molecular pathways of resistance.

The predictive models, trained on refined variant matrices, achieved reliable performance across multiple antibiotics, with gradient-boosted and ensemble classifiers outperforming simpler baselines. Similarly, lineage-informative mutations elucidated through PCA further suggested that resistance phenotypes are not uniformly distributed but tend to arise along specific evolutionary paths.

Nonetheless, several limitations remain. First, although variant effect annotation improved interpretability, the pipeline inherently depends on \textbf{SNP-based genomics}. Structural variants, mobile genetic elements, gene duplications, and plasmid-level dynamics, known drivers of AMR, remain underrepresented. Second, the current analysis is restricted to in silico predictions without experimental validation. Third, although these methods provided meaningful gene-level signals, they cannot fully replace mechanistic experiments; attention-based or attribution-based saliency is still correlational rather than causal. Finally, the approach does not yet incorporate temporal evolutionary signals beyond the PCA projection; resistance emergence is inherently dynamic, and richer longitudinal models could better capture adaptive trajectories.

\subsection{Future Work}
Future extensions of this work can significantly enhance both the biological depth and predictive capability of the AMR modeling framework. One important direction is the incorporation of structural variants and mobile genetic elements. While the current pipeline focuses on SNPs, many clinically relevant resistance mechanisms arise from \textbf{plasmid-borne genes}, \textbf{integrons}, \textbf{transposable elements}, and larger structural changes such as copy-number variation or gene duplications. Incorporating these signals, potentially through hybrid short and long read sequencing, would enable a more comprehensive representation of AMR determinants.

Another promising avenue is the \textbf{integration of pan-genome} and \textbf{gene presence/absence features}. Since the accessory genome often harbors key resistance genes, expanding the feature space beyond point mutations to include gene-level variability can provide a fuller view of the genomic landscape underlying resistance. Combining SNP effects with pan-genomic signatures would allow the model to simultaneously capture both fine-grained nucleotide changes and broader genomic architecture.

A third area of expansion is the incorporation of temporal evolutionary modeling. Although PCA-based analyses revealed distinct evolutionary trajectories, the current approach does not yet model the dynamics of adaptation explicitly. Temporal models, such as \textbf{phylogeny-aware predictors}, \textbf{recurrent neural networks}, or \textbf{Bayesian evolutionary frameworks}, would allow the prediction of future resistance states, the detection of early adaptive signals, and the inference of evolutionary paths under antibiotic pressure.

Finally, opportunities exist to optimize the pipeline for translational use and to broaden its multi-modal capabilities. Clinically viable AMR prediction requires faster variant calling, efficient annotation, and models that incorporate uncertainty quantification for decision support. Additionally, integrating other biological data types, such as \textbf{gene expression profiles} during antibiotic exposure, \textbf{proteomics} of efflux systems, or clinical metadata, could substantially improve both predictive performance and interpretability. Together, these directions lay a clear path toward a more comprehensive, mechanistically grounded, and clinically deployable AMR evolutionary modeling framework.

\printbibliography

@ARTICLE{Naylor2018-wz,
  title     = "Estimating the burden of antimicrobial resistance: a systematic
               literature review",
  author    = "Naylor, Nichola R and Atun, Rifat and Zhu, Nina and
               Kulasabanathan, Kavian and Silva, Sachin and Chatterjee, Anuja
               and Knight, Gwenan M and Robotham, Julie V",
  journal   = "Antimicrob. Resist. Infect. Control",
  publisher = "Springer Science and Business Media LLC",
  volume    =  7,
  number    =  1,
  pages     = "58",
  month     =  apr,
  year      =  2018,
  copyright = "https://creativecommons.org/licenses/by/4.0",
  language  = "en"
}

@ARTICLE{Boolchandani2019-cq,
  title     = "Sequencing-based methods and resources to study antimicrobial
               resistance",
  author    = "Boolchandani, Manish and D'Souza, Alaric W and Dantas, Gautam",
  journal   = "Nat. Rev. Genet.",
  publisher = "Springer Science and Business Media LLC",
  volume    =  20,
  number    =  6,
  pages     = "356--370",
  month     =  jun,
  year      =  2019,
  language  = "en"
}

@ARTICLE{Hutchings2019-du,
  title     = "Antibiotics: past, present and future",
  author    = "Hutchings, Matthew I and Truman, Andrew W and Wilkinson, Barrie",
  journal   = "Curr. Opin. Microbiol.",
  publisher = "Elsevier BV",
  volume    =  51,
  pages     = "72--80",
  month     =  oct,
  year      =  2019,
  copyright = "http://creativecommons.org/licenses/by/4.0/",
  language  = "en"
}

@ARTICLE{Salam2023-ye,
  title     = "Antimicrobial resistance: A growing serious threat for global
               public health",
  author    = "Salam, Md Abdus and Al-Amin, Md Yusuf and Salam, Moushumi
               Tabassoom and Pawar, Jogendra Singh and Akhter, Naseem and
               Rabaan, Ali A and Alqumber, Mohammed A A",
  journal   = "Healthcare (Basel)",
  publisher = "MDPI AG",
  volume    =  11,
  number    =  13,
  pages     = "1946",
  month     =  jul,
  year      =  2023,
  copyright = "https://creativecommons.org/licenses/by/4.0/",
  language  = "en"
}

@ARTICLE{Barnes2023-ov,
  title     = "Antimicrobial susceptibility testing to evaluate minimum
               inhibitory concentration values of clinically relevant
               antibiotics",
  author    = "Barnes, V, Lucien and Heithoff, Douglas M and Mahan, Scott P and
               House, John K and Mahan, Michael J",
  journal   = "STAR Protoc.",
  publisher = "Elsevier BV",
  volume    =  4,
  number    =  3,
  pages     = "102512",
  month     =  aug,
  year      =  2023,
  copyright = "http://creativecommons.org/licenses/by/4.0/",
  language  = "en"
}

@ARTICLE{Abushaheen2020-kr,
  title     = "Antimicrobial resistance, mechanisms and its clinical
               significance",
  author    = "Abushaheen, Manar Ali and {Muzaheed} and Fatani, Amal Jamil and
               Alosaimi, Mohammed and Mansy, Wael and George, Merin and
               Acharya, Sadananda and Rathod, Sanjay and Divakar, Darshan
               Devang and Jhugroo, Chitra and Vellappally, Sajith and Khan,
               Aftab Ahmed and Shaik, Jilani and Jhugroo, Poojdev",
  journal   = "Dis. Mon.",
  publisher = "Elsevier BV",
  volume    =  66,
  number    =  6,
  pages     = "100971",
  month     =  jun,
  year      =  2020,
  language  = "en"
}

@ARTICLE{Mustafa2024-bs,
  title     = "Whole genome sequencing: Applications in clinical bacteriology",
  author    = "Mustafa, Abu Salim",
  journal   = "Med. Princ. Pract.",
  publisher = "S. Karger AG",
  pages     = "1--13",
  month     =  feb,
  year      =  2024,
  copyright = "https://creativecommons.org/licenses/by-nc/4.0/",
  language  = "en"
}

@ARTICLE{Liu2020-hi,
  title    = "Evaluation of machine learning models for predicting
              antimicrobial resistance of Actinobacillus pleuropneumoniae from
              whole genome sequences",
  author   = "Liu, Zhichang and Deng, Dun and Lu, Huijie and Sun, Jian and Lv,
              Luchao and Li, Shuhong and Peng, Guanghui and Ma, Xianyong and
              Li, Jiazhou and Li, Zhenming and Rong, Ting and Wang, Gang",
  journal  = "Front. Microbiol.",
  volume   =  11,
  pages    = "48",
  month    =  feb,
  year     =  2020,
  language = "en"
}

@ARTICLE{Ren2022-ao,
  title     = "Prediction of antimicrobial resistance based on whole-genome
               sequencing and machine learning",
  author    = "Ren, Yunxiao and Chakraborty, Trinad and Doijad, Swapnil and
               Falgenhauer, Linda and Falgenhauer, Jane and Goesmann, Alexander
               and Hauschild, Anne-Christin and Schwengers, Oliver and Heider,
               Dominik",
  journal   = "Bioinformatics",
  publisher = "Oxford University Press (OUP)",
  volume    =  38,
  number    =  2,
  pages     = "325--334",
  month     =  jan,
  year      =  2022,
  copyright = "https://creativecommons.org/licenses/by-nc/4.0/",
  language  = "en"
}

@ARTICLE{Ren2022-uf,
  title     = "Multi-label classification for multi-drug resistance prediction
               of Escherichia coli",
  author    = "Ren, Yunxiao and Chakraborty, Trinad and Doijad, Swapnil and
               Falgenhauer, Linda and Falgenhauer, Jane and Goesmann, Alexander
               and Schwengers, Oliver and Heider, Dominik",
  journal   = "Comput. Struct. Biotechnol. J.",
  publisher = "Elsevier BV",
  volume    =  20,
  pages     = "1264--1270",
  month     =  mar,
  year      =  2022,
  copyright = "http://creativecommons.org/licenses/by/4.0/",
  language  = "en"
}

@ARTICLE{Sevillya2020-iz,
  title     = "Detecting horizontal gene transfer: a probabilistic approach",
  author    = "Sevillya, Gur and Adato, Orit and Snir, Sagi",
  journal   = "BMC Genomics",
  publisher = "Springer Science and Business Media LLC",
  volume    =  21,
  number    = "Suppl 1",
  pages     = "106",
  month     =  mar,
  year      =  2020,
  copyright = "https://creativecommons.org/licenses/by/4.0",
  language  = "en"
}

@ARTICLE{He2022-tm,
  title     = "Repulsion and attraction in searching: A hybrid algorithm based
               on gravitational kernel and vital few for cancer driver gene
               prediction",
  author    = "He, Zhihui and Lin, Yingqing and Wei, Runguo and Liu, Cheng and
               Jiang, Dazhi",
  journal   = "Comput. Biol. Med.",
  publisher = "Elsevier BV",
  volume    =  151,
  number    = "Pt A",
  pages     = "106236",
  month     =  dec,
  year      =  2022,
  language  = "en"
}

@ARTICLE{Yurtseven2023-vf,
  title    = "Machine learning and phylogenetic analysis allow for predicting
              antibiotic resistance in M. tuberculosis",
  author   = "Yurtseven, Alper and Buyanova, Sofia and Agrawal, Amay Ajaykumar
              and Bochkareva, Olga O and Kalinina, Olga V",
  journal  = "BMC Microbiol.",
  volume   =  23,
  number   =  1,
  pages    = "404",
  month    =  dec,
  year     =  2023,
  language = "en"
}

@ARTICLE{Mowlaei2023-uh,
  title    = "{FSF-GA}: A feature selection framework for phenotype prediction
              using genetic algorithms",
  author   = "Mowlaei, Mohammad Erfan and Shi, Xinghua",
  journal  = "Genes (Basel)",
  volume   =  14,
  number   =  5,
  month    =  may,
  year     =  2023,
  language = "en"
}

@ARTICLE{Feldgarden2022-pi,
  title    = "Curation of the {AMRFinderPlus} databases: applications,
              functionality and impact",
  author   = "Feldgarden, Michael and Brover, Vyacheslav and Fedorov, Boris and
              Haft, Daniel H and Prasad, Arjun B and Klimke, William",
  journal  = "Microb. Genom.",
  volume   =  8,
  number   =  6,
  month    =  jun,
  year     =  2022,
  language = "en"
}

@article{andrews2010fastqc,
  author = {Andrews, S.},
  title = {FastQC: a quality control tool for high throughput sequence data},
  journal = {},
  year = {2010},
  url = {http://www.bioinformatics.babraham.ac.uk/projects/fastqc},
}

@misc{li2013aligning,
      title={Aligning sequence reads, clone sequences and assembly contigs with BWA-MEM}, 
      author={Heng Li},
      year={2013},
      eprint={1303.3997},
      archivePrefix={arXiv},
      primaryClass={q-bio.GN}
}

@ARTICLE{Li2009-zq,
  title     = "Fast and accurate short read alignment with {Burrows-Wheeler}
               transform",
  author    = "Li, Heng and Durbin, Richard",
  journal   = "Bioinformatics",
  publisher = "Oxford University Press (OUP)",
  volume    =  25,
  number    =  14,
  pages     = "1754--1760",
  month     =  jul,
  year      =  2009,
  copyright = "http://creativecommons.org/licenses/by-nc/2.0/uk/",
  language  = "en"
}

@ARTICLE{Li2009-ou,
  title     = "The Sequence {Alignment/Map} format and {SAMtools}",
  author    = "Li, Heng and Handsaker, Bob and Wysoker, Alec and Fennell, Tim
               and Ruan, Jue and Homer, Nils and Marth, Gabor and Abecasis,
               Goncalo and Durbin, Richard and {1000 Genome Project Data
               Processing Subgroup}",
  journal   = "Bioinformatics",
  publisher = "Oxford University Press (OUP)",
  volume    =  25,
  number    =  16,
  pages     = "2078--2079",
  month     =  aug,
  year      =  2009,
  copyright = "http://creativecommons.org/licenses/by-nc/2.0/uk/",
  language  = "en"
}

@ARTICLE{Danecek2011-aw,
  title     = "The variant call format and {VCFtools}",
  author    = "Danecek, Petr and Auton, Adam and Abecasis, Goncalo and Albers,
               Cornelis A and Banks, Eric and DePristo, Mark A and Handsaker,
               Robert E and Lunter, Gerton and Marth, Gabor T and Sherry,
               Stephen T and McVean, Gilean and Durbin, Richard and {1000
               Genomes Project Analysis Group}",
  journal   = "Bioinformatics",
  publisher = "Oxford University Press (OUP)",
  volume    =  27,
  number    =  15,
  pages     = "2156--2158",
  month     =  aug,
  year      =  2011,
  copyright = "http://creativecommons.org/licenses/by-nc/2.5",
  language  = "en"
}

@ARTICLE{Hanson2022-gu,
  title     = "The mutagenic consequences of {DNA} methylation within and
               across generations",
  author    = "Hanson, Haley E and Liebl, Andrea L",
  journal   = "Epigenomes",
  publisher = "MDPI AG",
  volume    =  6,
  number    =  4,
  pages     = "33",
  month     =  oct,
  year      =  2022,
  copyright = "https://creativecommons.org/licenses/by/4.0/",
  language  = "en"
}

@ARTICLE{Van_Belkum1998-so,
  title     = "Short-sequence {DNA} repeats in prokaryotic genomes",
  author    = "van Belkum, A and Scherer, S and van Alphen, L and Verbrugh, H",
  journal   = "Microbiol. Mol. Biol. Rev.",
  publisher = "American Society for Microbiology",
  volume    =  62,
  number    =  2,
  pages     = "275--293",
  month     =  jun,
  year      =  1998,
  language  = "en"
}

@ARTICLE{Sun2019-us,
  title     = "Editorial: Horizontal gene transfer mediated bacterial
               antibiotic resistance",
  author    = "Sun, Dongchang and Jeannot, Katy and Xiao, Yonghong and Knapp,
               Charles W",
  journal   = "Front. Microbiol.",
  publisher = "Frontiers Media SA",
  volume    =  10,
  pages     = "1933",
  month     =  aug,
  year      =  2019,
  copyright = "https://creativecommons.org/licenses/by/4.0/",
  language  = "en"
}

@INBOOK{Van_der_Auwera2013-uq,
  title     = "From {FastQ} data to high-confidence variant calls: The genome
               analysis toolkit best practices pipeline",
  booktitle = "Current Protocols in Bioinformatics",
  author    = "Van der Auwera, Geraldine A and Carneiro, Mauricio O and Hartl,
               Christopher and Poplin, Ryan and del Angel, Guillermo and
               Levy-Moonshine, Ami and Jordan, Tadeusz and Shakir, Khalid and
               Roazen, David and Thibault, Joel and Banks, Eric and Garimella,
               Kiran V and Altshuler, David and Gabriel, Stacey and DePristo,
               Mark A",
  publisher = "John Wiley \& Sons, Inc.",
  pages     = "11.10.1--11.10.33",
  month     =  oct,
  year      =  2013,
  address   = "Hoboken, NJ, USA"
}

@INPROCEEDINGS{Zhao2018-yi,
  title           = "A study on optimizing {MarkDuplicate} in genome sequencing
                     pipeline",
  booktitle       = "Proceedings of the 2018 5th International Conference on
                     Bioinformatics Research and Applications",
  author          = "Zhao, Qi",
  publisher       = "ACM",
  month           =  dec,
  year            =  2018,
  address         = "New York, NY, USA",
  copyright       = "http://www.acm.org/publications/policies/copyright\_policy\#Background",
  conference      = "ICBRA '18: 2018 5th International Conference on
                     Bioinformatics Research and Applications",
  location        = "Hong Kong Hong Kong"


}

@ARTICLE{Muteeb2023-bl,
  title     = "Origin of antibiotics and antibiotic resistance, and their
               impacts on drug development: A narrative review",
  author    = "Muteeb, Ghazala and Rehman, Md Tabish and Shahwan, Moayad and
               Aatif, Mohammad",
  journal   = "Pharmaceuticals (Basel)",
  publisher = "MDPI AG",
  volume    =  16,
  number    =  11,
  pages     = "1615",
  month     =  nov,
  year      =  2023,
  copyright = "https://creativecommons.org/licenses/by/4.0/",
  language  = "en"
}

@ARTICLE{Dhingra2020-tq,
  title     = "Microbial resistance movements: An overview of global public
               health threats posed by antimicrobial resistance, and how best
               to counter",
  author    = "Dhingra, Sameer and Rahman, Nor Azlina A and Peile, Ed and
               Rahman, Motiur and Sartelli, Massimo and Hassali, Mohamed Azmi
               and Islam, Tariqul and Islam, Salequl and Haque, Mainul",
  journal   = "Front. Public Health",
  publisher = "Frontiers Media SA",
  volume    =  8,
  pages     = "535668",
  month     =  nov,
  year      =  2020,
  copyright = "https://creativecommons.org/licenses/by/4.0/",
  language  = "en"
}

@article{AHMED2024100081,
title = {Antimicrobial resistance: Impacts, challenges, and future prospects},
journal = {Journal of Medicine, Surgery, and Public Health},
volume = {2},
pages = {100081},
year = {2024},
issn = {2949-916X},
doi = {https://doi.org/10.1016/j.glmedi.2024.100081},
url = {https://www.sciencedirect.com/science/article/pii/S2949916X24000343},
author = {Sirwan Khalid Ahmed and Safin Hussein and Karzan Qurbani and Radhwan Hussein Ibrahim and Abdulmalik Fareeq and Kochr Ali Mahmood and Mona Gamal Mohamed}
}

@ARTICLE{Gajic2022-kl,
  title     = "Antimicrobial susceptibility testing: A comprehensive review of
               currently used methods",
  author    = "Gajic, Ina and Kabic, Jovana and Kekic, Dusan and Jovicevic,
               Milos and Milenkovic, Marina and Mitic Culafic, Dragana and
               Trudic, Anika and Ranin, Lazar and Opavski, Natasa",
  journal   = "Antibiotics (Basel)",
  publisher = "MDPI AG",
  volume    =  11,
  number    =  4,
  pages     = "427",
  month     =  mar,
  year      =  2022,
  copyright = "https://creativecommons.org/licenses/by/4.0/",
  language  = "en"
}

@ARTICLE{Alcock2023-gh,
  title     = "{CARD} 2023: expanded curation, support for machine learning,
               and resistome prediction at the Comprehensive Antibiotic
               Resistance Database",
  author    = "Alcock, Brian P and Huynh, William and Chalil, Romeo and Smith,
               Keaton W and Raphenya, Amogelang R and Wlodarski, Mateusz A and
               Edalatmand, Arman and Petkau, Aaron and Syed, Sohaib A and
               Tsang, Kara K and Baker, Sheridan J C and Dave, Mugdha and
               McCarthy, Madeline C and Mukiri, Karyn M and Nasir, Jalees A and
               Golbon, Bahar and Imtiaz, Hamna and Jiang, Xingjian and Kaur,
               Komal and Kwong, Megan and Liang, Zi Cheng and Niu, Keyu C and
               Shan, Prabakar and Yang, Jasmine Y J and Gray, Kristen L and
               Hoad, Gemma R and Jia, Baofeng and Bhando, Timsy and Carfrae,
               Lindsey A and Farha, Maya A and French, Shawn and Gordzevich,
               Rodion and Rachwalski, Kenneth and Tu, Megan M and Bordeleau,
               Emily and Dooley, Damion and Griffiths, Emma and Zubyk, Haley L
               and Brown, Eric D and Maguire, Finlay and Beiko, Robert G and
               Hsiao, William W L and Brinkman, Fiona S L and Van Domselaar,
               Gary and McArthur, Andrew G",
  journal   = "Nucleic Acids Res.",
  publisher = "Oxford University Press (OUP)",
  volume    =  51,
  number    = "D1",
  pages     = "D690--D699",
  month     =  jan,
  year      =  2023,
  copyright = "https://creativecommons.org/licenses/by/4.0/",
  language  = "en"
}

@ARTICLE{Kim2022-fi,
  title     = "Machine learning for antimicrobial resistance prediction:
               Current practice, limitations, and clinical perspective",
  author    = "Kim, Jee In and Maguire, Finlay and Tsang, Kara K and
               Gouliouris, Theodore and Peacock, Sharon J and McAllister, Tim A
               and McArthur, Andrew G and Beiko, Robert G",
  journal   = "Clin. Microbiol. Rev.",
  publisher = "American Society for Microbiology",
  volume    =  35,
  number    =  3,
  pages     = "e0017921",
  month     =  sep,
  year      =  2022,
  copyright = "https://doi.org/10.1128/ASMCopyrightv2",
  language  = "en"
}

@ARTICLE{Florensa2022-jf,
  title     = "{ResFinder} - an open online resource for identification of
               antimicrobial resistance genes in next-generation sequencing
               data and prediction of phenotypes from genotypes",
  author    = "Florensa, Alfred Ferrer and Kaas, Rolf Sommer and Clausen,
               Philip Thomas Lanken Conradsen and Aytan-Aktug, Derya and
               Aarestrup, Frank M",
  journal   = "Microb. Genom.",
  publisher = "Microbiology Society",
  volume    =  8,
  number    =  1,
  month     =  jan,
  year      =  2022,
  language  = "en"
}

@ARTICLE{Arango-Argoty2018-ej,
  title     = "{DeepARG}: a deep learning approach for predicting antibiotic
               resistance genes from metagenomic data",
  author    = "Arango-Argoty, Gustavo and Garner, Emily and Pruden, Amy and
               Heath, Lenwood S and Vikesland, Peter and Zhang, Liqing",
  journal   = "Microbiome",
  publisher = "Springer Science and Business Media LLC",
  volume    =  6,
  number    =  1,
  month     =  dec,
  year      =  2018,
  copyright = "http://creativecommons.org/licenses/by/4.0/",
  language  = "en"
}

@ARTICLE{Feldgarden2021-yl,
  title     = "{AMRFinderPlus} and the Reference Gene Catalog facilitate
               examination of the genomic links among antimicrobial resistance,
               stress response, and virulence",
  author    = "Feldgarden, Michael and Brover, Vyacheslav and
               Gonzalez-Escalona, Narjol and Frye, Jonathan G and Haendiges,
               Julie and Haft, Daniel H and Hoffmann, Maria and Pettengill,
               James B and Prasad, Arjun B and Tillman, Glenn E and Tyson,
               Gregory H and Klimke, William",
  journal   = "Sci. Rep.",
  publisher = "Springer Science and Business Media LLC",
  volume    =  11,
  number    =  1,
  pages     = "12728",
  month     =  jun,
  year      =  2021,
  copyright = "https://creativecommons.org/licenses/by/4.0",
  language  = "en"
}

@misc{chindelevitch2022applyingdatatechnologiescombat,
      title={Applying data technologies to combat AMR: current status, challenges, and opportunities on the way forward}, 
      author={Leonid Chindelevitch and Elita Jauneikaite and Nicole E. Wheeler and Kasim Allel and Bede Yaw Ansiri-Asafoakaa and Wireko A. Awuah and Denis C. Bauer and Stephan Beisken and Kara Fan and Gary Grant and Michael Graz and Yara Khalaf and Veranja Liyanapathirana and Carlos Montefusco-Pereira and Lawrence Mugisha and Atharv Naik and Sylvia Nanono and Anthony Nguyen and Timothy Rawson and Kessendri Reddy and Juliana M. Ruzante and Anneke Schmider and Roman Stocker and Leonhardt Unruh and Daniel Waruingi and Heather Graz and Maarten van Dongen},
      year={2022},
      eprint={2208.04683},
      archivePrefix={arXiv},
      primaryClass={cs.CY},
      url={https://arxiv.org/abs/2208.04683}, 
}

@ARTICLE{10.3389/fmicb.2019.01933,
  
AUTHOR={Sun, Dongchang  and Jeannot, Katy  and Xiao, Yonghong  and Knapp, Charles W. },
         
TITLE={Editorial: Horizontal Gene Transfer Mediated Bacterial Antibiotic Resistance},
        
JOURNAL={Frontiers in Microbiology},
        
VOLUME={Volume 10 - 2019},

YEAR={2019},

URL={https://www.frontiersin.org/journals/microbiology/articles/10.3389/fmicb.2019.01933},

DOI={10.3389/fmicb.2019.01933},

ISSN={1664-302X},

}

@ARTICLE{Sun2023-bo,
  title    = "Editorial: Horizontal gene transfer mediated bacterial antibiotic
              resistance, volume {II}",
  author   = "Sun, Dongchang and Sun, Xingmin and Hu, Yongfei and Yamaichi,
              Yoshiharu",
  journal  = "Front. Microbiol.",
  volume   =  14,
  pages    = "1221606",
  month    =  jun,
  year     =  2023,
  language = "en"
}

@ARTICLE{Baquero2021-zm,
  title     = "Evolutionary pathways and trajectories in antibiotic resistance",
  author    = "Baquero, F and Mart{\'\i}nez, J L and F Lanza, V and
               Rodr{\'\i}guez-Beltr{\'a}n, J and Gal{\'a}n, J C and San
               Mill{\'a}n, A and Cant{\'o}n, R and Coque, T M",
  journal   = "Clin. Microbiol. Rev.",
  publisher = "American Society for Microbiology",
  volume    =  34,
  number    =  4,
  pages     = "e0005019",
  month     =  dec,
  year      =  2021,
  copyright = "https://doi.org/10.1128/ASMCopyrightv2",
  language  = "en"
}

@ARTICLE{Wilson2016-ap,
  title     = "The population genetics of drug resistance evolution in natural
               populations of viral, bacterial and eukaryotic pathogens",
  author    = "Wilson, Benjamin A and Garud, Nandita R and Feder, Alison F and
               Assaf, Zoe J and Pennings, Pleuni S",
  journal   = "Mol. Ecol.",
  publisher = "Wiley",
  volume    =  25,
  number    =  1,
  pages     = "42--66",
  month     =  jan,
  year      =  2016,
  copyright = "http://onlinelibrary.wiley.com/termsAndConditions\#vor",
  language  = "en"
}

@ARTICLE{Hasan2021-pm,
  title     = "Revisiting antibiotic resistance: Mechanistic foundations to
               evolutionary outlook",
  author    = "Hasan, Chowdhury M and Dutta, Debprasad and Nguyen, An N T",
  journal   = "Antibiotics (Basel)",
  publisher = "MDPI AG",
  volume    =  11,
  number    =  1,
  pages     = "40",
  month     =  dec,
  year      =  2021,
  copyright = "https://creativecommons.org/licenses/by/4.0/",
  language  = "en"
}

@ARTICLE{Gupta2014-ue,
  title     = "{ARG-ANNOT}, a new bioinformatic tool to discover antibiotic
               resistance genes in bacterial genomes",
  author    = "Gupta, Sushim Kumar and Padmanabhan, Babu Roshan and Diene,
               Seydina M and Lopez-Rojas, Rafael and Kempf, Marie and Landraud,
               Luce and Rolain, Jean-Marc",
  journal   = "Antimicrob. Agents Chemother.",
  publisher = "American Society for Microbiology",
  volume    =  58,
  number    =  1,
  pages     = "212--220",
  year      =  2014,
  language  = "en"
}

@ARTICLE{Hu2024-qx,
  title     = "Assessing computational predictions of antimicrobial resistance
               phenotypes from microbial genomes",
  author    = "Hu, Kaixin and Meyer, Fernando and Deng, Zhi-Luo and Asgari,
               Ehsaneddin and Kuo, Tzu-Hao and M{\"u}nch, Philipp C and
               McHardy, Alice C",
  journal   = "Brief. Bioinform.",
  publisher = "Oxford University Press (OUP)",
  volume    =  25,
  number    =  3,
  month     =  mar,
  year      =  2024,
  copyright = "https://creativecommons.org/licenses/by-nc/4.0/",
  language  = "en"
}

@ARTICLE{Valavarasu2025-js,
  title     = "Prediction of antibiotic resistance from antibiotic
               susceptibility testing results from surveillance data using
               machine learning",
  author    = "Valavarasu, Swetha and Sangu, Yasaswini and Mahapatra, Tanmaya",
  journal   = "Sci. Rep.",
  publisher = "Springer Science and Business Media LLC",
  volume    =  15,
  number    =  1,
  pages     = "30509",
  month     =  aug,
  year      =  2025,
  copyright = "https://creativecommons.org/licenses/by/4.0",
  language  = "en"
}

@ARTICLE{Rayesha2025-aa,
  title     = "Antibiotic genomic resistance prediction using deep learning
               models",
  author    = "Rayesha, S M Shifana and Banu, W Aisha and Rahman, Afzalur",
  journal   = "Int. J. Bioinform. Res. Appl.",
  publisher = "Inderscience Publishers",
  volume    =  21,
  number    =  2,
  pages     = "121--136",
  year      =  2025,
  language  = "en"
}

@INPROCEEDINGS{10957126,
  author={Preethi, R and Bharati, Ritul and Priya, S.},
  booktitle={2024 Third International Conference on Artificial Intelligence, Computational Electronics and Communication System (AICECS)}, 
  title={Predicting Antibiotic Resistance from Genomic Sequences Using a Hybrid CNN-RNN Model: A Comprehensive Approach}, 
  year={2024},
  volume={},
  number={},
  pages={1-6},
  doi={10.1109/AICECS63354.2024.10957126}
}

@ARTICLE{Lakin2019-lp,
  title     = "Hierarchical Hidden Markov models enable accurate and diverse
               detection of antimicrobial resistance sequences",
  author    = "Lakin, Steven M and Kuhnle, Alan and Alipanahi, Bahar and Noyes,
               Noelle R and Dean, Chris and Muggli, Martin and Raymond, Rob and
               Abdo, Zaid and Prosperi, Mattia and Belk, Keith E and Morley,
               Paul S and Boucher, Christina",
  journal   = "Commun. Biol.",
  publisher = "Springer Science and Business Media LLC",
  volume    =  2,
  number    =  1,
  pages     = "294",
  month     =  aug,
  year      =  2019,
  copyright = "https://creativecommons.org/licenses/by/4.0",
  language  = "en"
}

@ARTICLE{Rowe2018-kn,
  title     = "Indexed variation graphs for efficient and accurate resistome
               profiling",
  author    = "Rowe, Will P M and Winn, Martyn D",
  journal   = "Bioinformatics",
  publisher = "Oxford University Press (OUP)",
  volume    =  34,
  number    =  21,
  pages     = "3601--3608",
  month     =  nov,
  year      =  2018,
  copyright = "http://creativecommons.org/licenses/by/4.0/",
  language  = "en"
}

@ARTICLE{Von_Wintersdorff2016-dk,
  title     = "Dissemination of antimicrobial resistance in microbial
               ecosystems through horizontal gene transfer",
  author    = "von Wintersdorff, Christian J H and Penders, John and van
               Niekerk, Julius M and Mills, Nathan D and Majumder, Snehali and
               van Alphen, Lieke B and Savelkoul, Paul H M and Wolffs, Petra F
               G",
  journal   = "Front. Microbiol.",
  publisher = "Frontiers Media SA",
  volume    =  7,
  pages     = "173",
  month     =  feb,
  year      =  2016,
  language  = "en"
}

@ARTICLE{Bottery2021-px,
  title     = "Ecology and evolution of antimicrobial resistance in bacterial
               communities",
  author    = "Bottery, Michael J and Pitchford, Jonathan W and Friman,
               Ville-Petri",
  journal   = "ISME J.",
  publisher = "Oxford University Press (OUP)",
  volume    =  15,
  number    =  4,
  pages     = "939--948",
  month     =  apr,
  year      =  2021,
  copyright = "https://creativecommons.org/licenses/by/4.0/",
  language  = "en"
}

@misc{alexandre2024bridgingwrightfishermoranmodels,
      title={Bridging Wright-Fisher and Moran models}, 
      author={Arthur Alexandre and Alia Abbara and Cecilia Fruet and Claude Loverdo and Anne-Florence Bitbol},
      year={2024},
      eprint={2407.12560},
      archivePrefix={arXiv},
      primaryClass={q-bio.PE},
      doi={https://doi.org/10.1016/j.jtbi.2024.112030},
      url={https://arxiv.org/abs/2407.12560}, 
}

@ARTICLE{Das2020-zh,
  title     = "Predictable properties of fitness landscapes induced by
               adaptational tradeoffs",
  author    = "Das, Suman G and Direito, Susana Ol and Waclaw, Bartlomiej and
               Allen, Rosalind J and Krug, Joachim",
  journal   = "Elife",
  publisher = "eLife Sciences Publications, Ltd",
  volume    =  9,
  number    = "e55155",
  month     =  may,
  year      =  2020,
  copyright = "http://creativecommons.org/licenses/by/4.0/",
  language  = "en"
}

@ARTICLE{Harmand2017-vb,
  title     = "Fisher's geometrical model and the mutational patterns of
               antibiotic resistance across dose gradients",
  author    = "Harmand, No{\'e}mie and Gallet, Romain and Jabbour-Zahab, Roula
               and Martin, Guillaume and Lenormand, Thomas",
  journal   = "Evolution",
  publisher = "Oxford University Press (OUP)",
  volume    =  71,
  number    =  1,
  pages     = "23--37",
  month     =  jan,
  year      =  2017,
  copyright = "http://onlinelibrary.wiley.com/termsAndConditions",
  language  = "en"
}

@ARTICLE{Chait2017-qy,
  title     = "Shaping bacterial population behavior through
               computer-interfaced control of individual cells",
  author    = "Chait, Remy and Ruess, Jakob and Bergmiller, Tobias and Tka{\v
               c}ik, Ga{\v s}per and Guet, C{\u a}lin C",
  journal   = "Nat. Commun.",
  publisher = "Springer Science and Business Media LLC",
  volume    =  8,
  number    =  1,
  pages     = "1535",
  month     =  nov,
  year      =  2017,
  copyright = "https://creativecommons.org/licenses/by/4.0",
  language  = "en"
}

@ARTICLE{Ragalo2018-sk,
  title     = "Evolving dynamic fitness measures for genetic programming",
  author    = "Ragalo, Anisa and Pillay, Nelishia",
  journal   = "Expert Syst. Appl.",
  publisher = "Elsevier BV",
  volume    =  109,
  pages     = "162--187",
  month     =  nov,
  year      =  2018,
  language  = "en"
}

@ARTICLE{Colin2020-zx,
  title     = "Genetic algorithms as a tool for dosing guideline optimization:
               Application to intermittent infusion dosing for vancomycin in
               adults",
  author    = "Colin, Pieter J and Eleveld, Douglas J and Thomson, Alison H",
  journal   = "CPT Pharmacometrics Syst. Pharmacol.",
  publisher = "Wiley",
  volume    =  9,
  number    =  5,
  pages     = "294--302",
  month     =  may,
  year      =  2020,
  copyright = "http://creativecommons.org/licenses/by-nc-nd/4.0/",
  language  = "en"
}

@ARTICLE{Yurtseven2023-lf,
  title    = "Machine learning and phylogenetic analysis allow for predicting
              antibiotic resistance in M. tuberculosis",
  author   = "Yurtseven, Alper and Buyanova, Sofia and Agrawal, Amay Ajaykumar
              and Bochkareva, Olga O and Kalinina, Olga V",
  journal  = "BMC Microbiol.",
  volume   =  23,
  number   =  1,
  pages    = "404",
  month    =  dec,
  year     =  2023,
  language = "en"
}

@ARTICLE{Brito2021-tp,
  title     = "Examining horizontal gene transfer in microbial communities",
  author    = "Brito, Ilana Lauren",
  journal   = "Nat. Rev. Microbiol.",
  publisher = "Springer Science and Business Media LLC",
  volume    =  19,
  number    =  7,
  pages     = "442--453",
  month     =  jul,
  year      =  2021,
  language  = "en"
}

@ARTICLE{Brown2016-ct,
  title     = "{SimBac}: simulation of whole bacterial genomes with homologous
               recombination",
  author    = "Brown, Thomas and Didelot, Xavier and Wilson, Daniel J and Maio,
               Nicola De",
  journal   = "Microb. Genom.",
  publisher = "Microbiology Society",
  volume    =  2,
  number    =  1,
  month     =  jan,
  year      =  2016,
  language  = "en"
}

@ARTICLE{Didelot2007-tv,
  title     = "Inference of bacterial microevolution using multilocus sequence
               data",
  author    = "Didelot, Xavier and Falush, Daniel",
  journal   = "Genetics",
  publisher = "Oxford University Press (OUP)",
  volume    =  175,
  number    =  3,
  pages     = "1251--1266",
  month     =  mar,
  year      =  2007,
  copyright = "https://academic.oup.com/journals/pages/open\_access/funder\_policies/chorus/standard\_publication\_model",
  language  = "en"
}

@ARTICLE{Wang2024-zm,
  title     = "Inter-plasmid transfer of antibiotic resistance genes
               accelerates antibiotic resistance in bacterial pathogens",
  author    = "Wang, Xiaolong and Zhang, Hanhui and Yu, Shenbo and Li, Donghang
               and Gillings, Michael R and Ren, Hongqiang and Mao, Daqing and
               Guo, Jianhua and Luo, Yi",
  journal   = "ISME J.",
  publisher = "Oxford University Press (OUP)",
  volume    =  18,
  number    =  1,
  month     =  jan,
  year      =  2024,
  copyright = "https://creativecommons.org/licenses/by/4.0/",
  language  = "en"
}

@ARTICLE{Vrancianu2020-jq,
  title     = "Targeting plasmids to limit acquisition and transmission of
               antimicrobial resistance",
  author    = "Vrancianu, Corneliu Ovidiu and Popa, Laura Ioana and Bleotu,
               Coralia and Chifiriuc, Mariana Carmen",
  journal   = "Front. Microbiol.",
  publisher = "Frontiers Media SA",
  volume    =  11,
  pages     = "761",
  month     =  may,
  year      =  2020,
  copyright = "https://creativecommons.org/licenses/by/4.0/",
  language  = "en"
}

@ARTICLE{Jacobs1991-ir,
  title     = "Adaptive mixtures of local experts",
  author    = "Jacobs, Robert A and Jordan, Michael I and Nowlan, Steven J and
               Hinton, Geoffrey E",
  journal   = "Neural Comput.",
  publisher = "MIT Press - Journals",
  volume    =  3,
  number    =  1,
  pages     = "79--87",
  year      =  1991,
  language  = "en"
}

@article{Minoura_etal_2021_scMM,
  author    = {Minoura, Kazuyuki and others},
  title     = {A mixture-of-experts deep generative model for integrated single-cell multi-omics data},
  journal   = {Bioinformatics},
  year      = {2021},
  volume    = {37},
  number    = {Supplement 1},
  pages     = {i116–i124},
  doi       = {10.1093/bioinformatics/btab265},
}

@article{Sun_etal_2024_AIDOProtein,
  author    = {Sun, Ning and Zou, Shuxian and Tao, Tianhua and Mahbub, Sazan and Li, Dian and Zhuang, Yonghao and … Xing, Eric P.},
  title     = {Mixture of Experts Enable Efficient and Effective Protein Understanding and Design},
  journal   = {bioRxiv preprint},
  year      = {2024},
  doi       = {10.1101/2024.11.29.625425},
}

@ARTICLE{Spooner2025-tt,
  title    = "Benchmarking ensemble machine learning algorithms for
              multi-class, multi-omics data integration in clinical outcome
              prediction",
  author   = "Spooner, Annette and Moridani, Mohammad Karimi and Toplis, Barbra
              and Behary, Jason and Safarchi, Azadeh and Maher, Salim and
              Vafaee, Fatemeh and Zekry, Amany and Sowmya, Arcot",
  journal  = "Brief. Bioinform.",
  volume   =  26,
  number   =  2,
  month    =  mar,
  year     =  2025,
  language = "en"
}

@ARTICLE{Guan2022-js,
  title     = "Classification of protein sequences by a novel alignment-free
               method on bacterial and virus families",
  author    = "Guan, Mengcen and Zhao, Leqi and Yau, Stephen S-T",
  journal   = "Genes (Basel)",
  publisher = "MDPI AG",
  volume    =  13,
  number    =  10,
  pages     = "1744",
  month     =  sep,
  year      =  2022,
  copyright = "https://creativecommons.org/licenses/by/4.0/",
  language  = "en"
}

@ARTICLE{Stanley2002-lk,
  title     = "Evolving neural networks through augmenting topologies",
  author    = "Stanley, Kenneth O and Miikkulainen, Risto",
  journal   = "Evol. Comput.",
  publisher = "MIT Press - Journals",
  volume    =  10,
  number    =  2,
  pages     = "99--127",
  year      =  2002,
  language  = "en"
}

@misc{miikkulainen2017evolvingdeepneuralnetworks,
      title={Evolving Deep Neural Networks}, 
      author={Risto Miikkulainen and Jason Liang and Elliot Meyerson and Aditya Rawal and Dan Fink and Olivier Francon and Bala Raju and Hormoz Shahrzad and Arshak Navruzyan and Nigel Duffy and Babak Hodjat},
      year={2017},
      eprint={1703.00548},
      archivePrefix={arXiv},
      primaryClass={cs.NE},
      url={https://arxiv.org/abs/1703.00548}, 
}

@misc{khadka2018evolutionguidedpolicygradientreinforcement,
      title={Evolution-Guided Policy Gradient in Reinforcement Learning}, 
      author={Shauharda Khadka and Kagan Tumer},
      year={2018},
      eprint={1805.07917},
      archivePrefix={arXiv},
      primaryClass={cs.LG},
      url={https://arxiv.org/abs/1805.07917}, 
}

@ARTICLE{Riou2023-uw,
  title    = "Projecting the development of antimicrobial resistance in
              Neisseria gonorrhoeae from antimicrobial surveillance data: a
              mathematical modelling study",
  author   = "Riou, Julien and Althaus, Christian L and Allen, Hester and Cole,
              Michelle J and Grad, Yonatan H and Heijne, Janneke C M and Unemo,
              Magnus and Low, Nicola",
  journal  = "BMC Infect. Dis.",
  volume   =  23,
  number   =  1,
  pages    = "252",
  month    =  apr,
  year     =  2023,
  language = "en"
}

@ARTICLE{Dong2025-rt,
  title     = "Painting peptides with antimicrobial potency through deep
               reinforcement learning",
  author    = "Dong, Ruihan and Cao, Qiushi and Song, Chen",
  journal   = "Adv. Sci. (Weinh.)",
  publisher = "Wiley",
  number    = "e06332",
  pages     = "e06332",
  month     =  sep,
  year      =  2025,
  copyright = "http://creativecommons.org/licenses/by/4.0/",
  language  = "en"
}

@ARTICLE{Needleman1970-yy,
  title     = "A general method applicable to the search for similarities in
               the amino acid sequence of two proteins",
  author    = "Needleman, S B and Wunsch, C D",
  journal   = "J. Mol. Biol.",
  publisher = "Elsevier BV",
  volume    =  48,
  number    =  3,
  pages     = "443--453",
  month     =  mar,
  year      =  1970,
  language  = "en"
}

\end{document}